\documentclass[fleqn,10pt]{wlscirep}
\usepackage[utf8]{inputenc}
\usepackage[T1]{fontenc}
\usepackage{units}
\usepackage{esvect}
\usepackage{subfigure}
\usepackage{newtxtext,newtxmath}
\usepackage{lineno}
\usepackage[title]{appendix}
\newcommand*\patchAmsMathEnvironmentForLineno[1]{
  \expandafter\let\csname old#1\expandafter\endcsname\csname #1\endcsname
  \expandafter\let\csname oldend#1\expandafter\endcsname\csname end#1\endcsname
  \renewenvironment{#1}
     {\linenomath\csname old#1\endcsname}
     {\csname oldend#1\endcsname\endlinenomath}}
\newcommand*\patchBothAmsMathEnvironmentsForLineno[1]{
  \patchAmsMathEnvironmentForLineno{#1}
  \patchAmsMathEnvironmentForLineno{#1*}}
\AtBeginDocument{
\patchBothAmsMathEnvironmentsForLineno{equation}
\patchBothAmsMathEnvironmentsForLineno{align}
\patchBothAmsMathEnvironmentsForLineno{flalign}
\patchBothAmsMathEnvironmentsForLineno{alignat}
\patchBothAmsMathEnvironmentsForLineno{gather}
\patchBothAmsMathEnvironmentsForLineno{multline}
}

\DeclareMathOperator{\E}{\mathbb{E}}
\renewcommand{\d}{\mathrm{d}}
\newcommand{\define}{\stackrel{\text{def}}{=}}
\newcommand{\degree}{\ensuremath{^{\circ}}}
\title{Estimating the population mean for 
a vertical profile of energy dissipation rate}

\author[1,*]{Nozomi Sugiura}
\author[1]{Shinya Kouketsu}
\author[1]{Shuhei Masuda}
\author[1]{Satoshi Osafune}
\author[2]{Ichiro Yasuda}

\affil[1]{Research and Development Center for Global Change, Japan Agency for Marine-Earth Science and Technology, Yokosuka, Japan}
\affil[2]{Atmosphere and Ocean Research Institute, University of Tokyo, Chiba, Japan}

\affil[*]{nsugiura@jamstec.go.jp}

\begin{abstract}
Energy dissipation rates are an important characteristic of turbulence; however, their magnitude in observational profiles can be incorrectly determined owing to their irregular appearance during vertical evolution. By analysing the data obtained from oceanic turbulence measurements, we demonstrate that the vertical sequences of energy dissipation rates exhibit a scaling property. Utilising this property, we propose a method to estimate the population mean for a profile. For scaling in the observed profiles, we demonstrate that our data exhibit a statistical property consistent with that exhibited by the universal multifractal model. Meanwhile, the population mean and its uncertainty can be estimated by inverting the probability distribution obtained by Monte Carlo simulations of a cascade model; to this end, observational constraints from several moments are imposed over each vertical sequence. This approach enables us to determine, to some extent, whether a profile shows an occasionally large mean or whether the population mean itself is large. Thus, it will contribute to the refinement of the regional estimation of the ocean energy budget, where only a small amount of turbulence observation data is available.
\end{abstract}
\begin{document}
\flushbottom
\maketitle
\thispagestyle{fancy}

\section*{Introduction} \label{Intro}
Numerous existing studies have highlighted the importance of determining energy dissipation rates
to investigate ocean general circulation 
\cite{gregg1973vertical,munk1998abyssal}. 
Therefore, 
several observational studies have been conducted to obtain the vertical profiles of
energy dissipation rates using ocean microstructure profilers \cite{Waterhouse2014,ucsddatabase}.
In addition, to understand the statistics of the irregular evolution of observational profiles, 
studies have been conducted 
from the perspective of statistical fluid mechanics, 
as summarised below.

In fully developed turbulence,
an inertial subrange of length scales exists wherein 
the advective term dominates the molecular viscosity term 
in the Navier--Stokes equation
\cite{pope2000turbulent}.
In this inertial subrange, 
a cascade of energies can be observed from large to small scales,
as intuitively stated by Richardson\cite{richardson2007weather}.
In the first quantitative theory on energy cascades,
Kolmogorov\cite{kolmogorov1941local} established a relationship wherein velocity fluctuations are locally isotropic and 
are determined by the homogeneous
energy dissipation rate; here, homogeneous means that the statistical property is independent of the position $x$, 
\begin{align}
  \left< |v(x+\ell)-v(x)| \right> &\approx \varepsilon^{1/3}\ell^{1/3},
\end{align}
where $v$ denotes the velocity; $\varepsilon$, 
the energy dissipation rate; $\ell$, the distance between the points; and 
$\left<\cdot\right>$, the expected value.
Subsequently, the energy dissipation rate
was argued to vary, exhibiting considerable random fluctuations
\cite{landau1987fluid}.
Thus, a refined theory\cite{kolmogorov1962refinement}
 was proposed to address this issue.
This theory stated that i) 
$\log{\varepsilon_r}$, which is the logarithm of the spatially averaged 
energy dissipation rate over scale $r$,
obeys a Gaussian distribution, and 
ii) its variance obeys $\sigma^2_{\log{\varepsilon_r}}
=A+\mu\log{(L/r)}$,
where $L$ denotes the outer scale;
$A$, a constant associated with the macrostructure of flow;
and $\mu$, the intermittency constant.

In addition, several experimental studies 
\cite{gurvich1963experimental,Pond1965} demonstrated 
that small-scale dissipation is a random field 
that has a spatial structure with power--law correlations,
\begin{align}
  \left<\varepsilon(x)\varepsilon(x+\ell)\right>&\propto \ell^{-\mu},\quad \ell>0.
  \label{corr}
\end{align}

Then, Yaglom\cite{Yaglom1966} formulated a quantitative 
model, which was consistent with the log-normal scaling
presented by Kolmogorov\cite{kolmogorov1962refinement}
and the power--law correlations, 
as a multiplicative cascade, 
where $\varepsilon_{r}$ was expressed with a binary tree 
comprising independent and identically distributed (i.i.d) 
random variables, $W_{n',k}$ ($\sim W$),
\begin{align}
  \forall~ 1 \leq j \leq 2^n, \quad \varepsilon_r(x_j)&=
  \prod_{n'=1}^{n} W_{n',\lfloor (j-1)/2^{n-n'} \rfloor+1},\label{vertS}
\end{align}
where $x_j$ are the positions with equal spacing and
$\lfloor s \rfloor$ is the floor function, which assigns the integer that 
satisfies $0 \leq s-\lfloor s \rfloor<1$.
If the random variables are set to have the moment exponent
$K(q)=\log_{2}\left<W^q\right>=(\mu/2)(q^2-q)$,
then the energy conservation in a probabilistic sense, $\left< W \right>=1$,
and the log-normal scaling in Kolmogorov (1962)\cite{kolmogorov1962refinement}
are reproduced.
Moreover, correlation (\ref{corr}) is reproduced 
because we have
$ \left<\varepsilon(x)\varepsilon(x+\ell)\right>=
\left< W^2 \right>^{n-m} \left< W \right>^{2m}
\propto \ell^{-K(2)}$, where $L=2^n r,~\ell = 2^m r$ for small $r$
\cite{Yaglom1966,monin2013statistical8}.

Several alternative %
multiplicative cascade models have been developed
with different generators, including the
$\beta$ model \cite{frisch1978simple},
random $\beta$ model \cite{benzi1984multifractal},
$\alpha$ model \cite{schertzer1984elliptical},
$p$ model \cite{meneveau1987simple},
log-stable model \cite{SchertzerLovejoy1987},
and log-Poisson model \cite{she1994universal}.
An important observation regarding Yaglom's cascade  
is that the property required for the law of random variable $W$
can be formulated such that the product of several random variables still obeys 
the same class of distribution, 
$\prod_{n'=1}^n W_{n'} \sim a_n W^{b_n}$, with $a_n,b_n>0$ 
\cite{schmitt_huang_2016}.
Consistent with this condition,
the universal multifractal model \cite{SchertzerLovejoy1987}
employs a stable L\'{e}vy generator, $\Gamma$, 
that is maximally left skewed and satisfies $W=\mathrm{e}^{\Gamma}$. 
This results in a simple and nonanalytic form of 
the moment exponent,
$K(q)=\left(C_1/(\alpha-1)\right)\left(q^{\alpha}-q\right),$
where $\alpha$ is the multifractal index, which can be a non-integer,
 and $C_1$ is the codimension of the mean.
The universal multifractal model %
is the most promising model.
This model can well reproduce the variability in 
several phenomena including turbulence, other geophysical phenomena,
and several fractal-like 
appearances in natural and man-made objects.

Based on this theory, 
we discuss a refined statistical treatment of 
the vertical profiles of the observed energy dissipation rates.
We first distinguish the 'mean energy dissipation rate',
which refers to the sample (arithmetic) mean over a profile,
and the 'energy input rate', which refers to the population mean for a profile.

Thus, we reconsider one of the basic questions in 
the observational study of ocean turbulence:
When a vertical profile of the energy dissipation rate is given,
how can one estimate the energy input rate or
the population mean of the energy dissipation rate for a profile, 
which has been commonly equated with 
the arithmetic mean over the profile?
Our question pertains to whether one can obtain 
information regarding the energy input rate
beyond the arithmetic mean.
The answer is yes, because we can construct a model 
for the turbulent cascade process and 
solve the inversion problem to obtain the energy input rate 
under an observational constraint.
In this study, we first 
show that the observed profiles of the depth-averaged energy dissipation rate, $\epsilon_r$, 
exhibit a scaling property consistent with that of the universal multifractal model. 
Then, we construct a multiplicative cascade simulation 
model that describes the statistics of the observational data.
Finally, we propose a method to explain certain statistics of the observed profiles 
based on the simulation model and develop an inversion method to 
estimate the energy input rate.
This result illustrates a systematic 
method of gaining further
quantitative information from profile data.

\section*{Methods}
\subsection*{Observational data \label{obs_data}}
\begin{figure}[ht]
  \begin{center}
    \includegraphics[width=0.7\columnwidth]{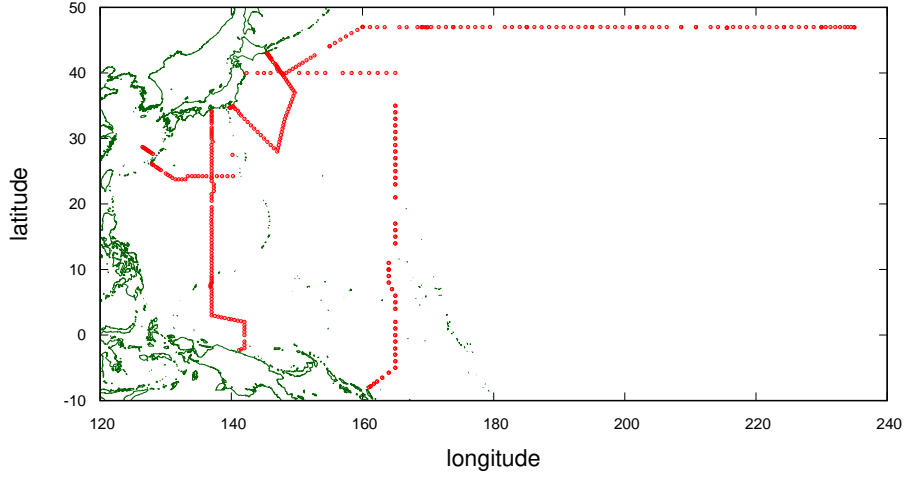}\\
    \caption{Horizontal locations of the observed profiles (red) and
      land--sea boundaries (green). 
      The units of longitude and latitude are $\degree \mathrm{E}$ and
      $\degree \mathrm{N}$, respectively.
      \label{map}}
  \end{center}
\end{figure}

In this section,
we describe the turbulence observational data employed in this study.
The data 
were retrieved from the Pacific Ocean (Fig.\,\ref{map}) \cite{goto2018}.
They comprise $I=409$ profiles, 
each of which typically extends over a depth of
$2000$ to $6000$ $\unit{m}$ below the sea surface, in turn
comprising observational bins with width of $r_0\simeq 10 \unit{m}$.
The turbulent energy dissipation rate for each bin, $\epsilon_{r_0}$, 
is derived by averaging the observational values in the bin,
which are estimated from the 
observed spectrum of the temperature vertical gradient
based on the procedure presented in Goto et al. (2016,2018)\cite{Goto2016,goto2018}
(see Supplementary information\,A for the estimation procedure).
We restrict our investigation to 
the intermittency occurring at larger scales, $r \geq r_0$.

\begin{figure}
  \begin{center}
    \subfigure[Original profiles $\epsilon_{r_0}$ in linear scale]
              {\includegraphics[width=0.45\columnwidth]{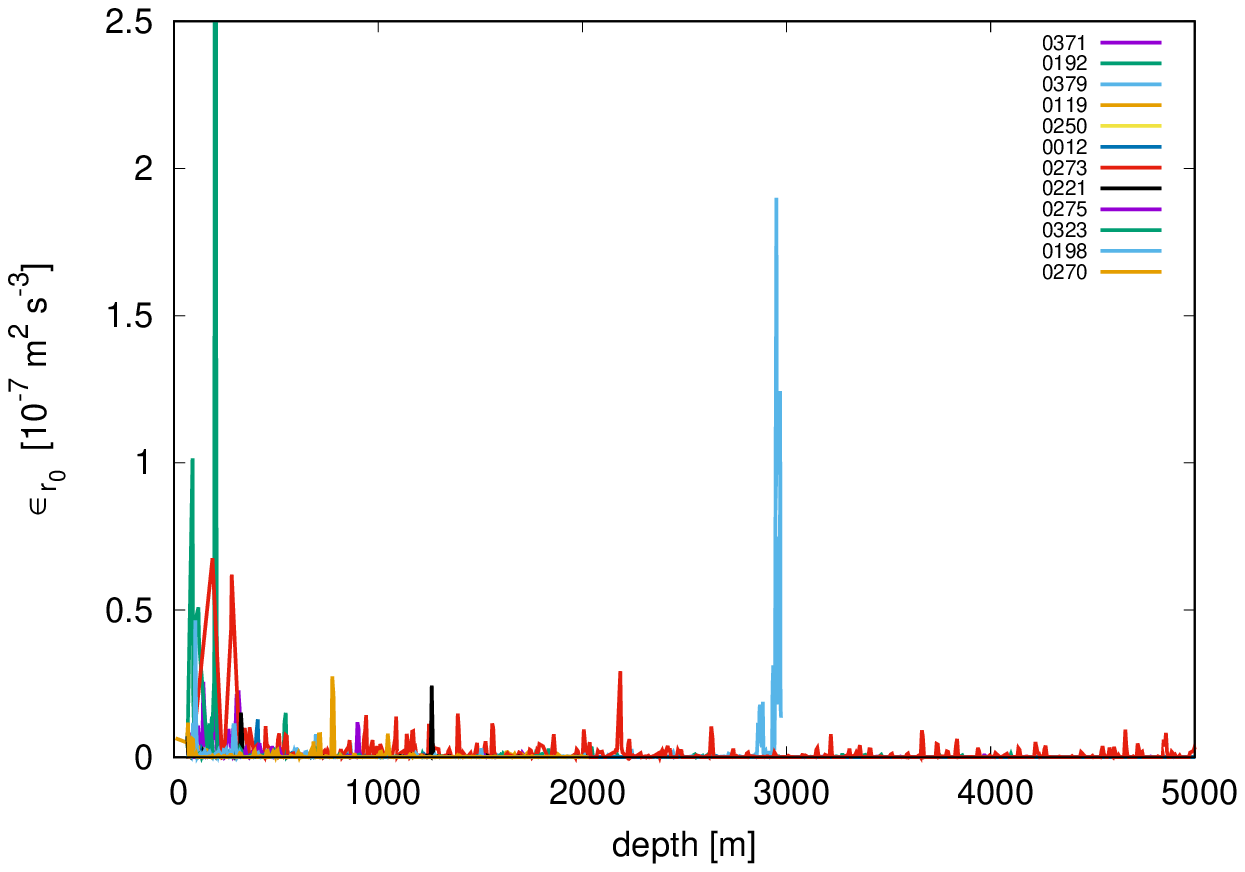}\label{depth-ave_a}}
              \subfigure[Original profiles $\log{\epsilon_{r_0}}$]
                        {\includegraphics[width=0.45\columnwidth]{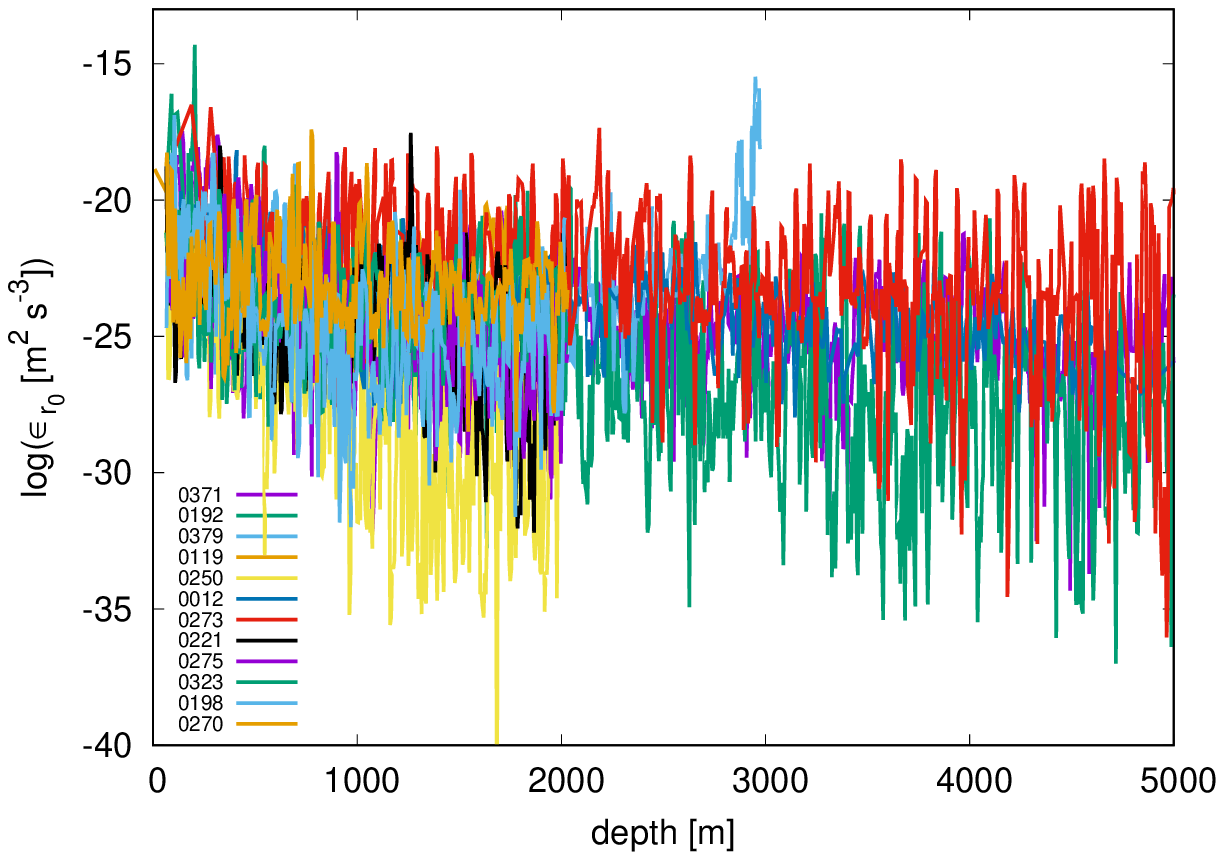}\label{depth-ave_b}}
                        \caption{Appearances of observed profiles.
                          \label{depth-ave}}
  \end{center}
\end{figure}

Let $r_0$ be the bin width, 
$\vv{x}_i$ the horizontal coordinate of the $i$-th profile,
and $z^i_j$ the vertical coordinate of the $j$-th point
in the $i$-th profile.
These positive-valued data
exhibit the following characteristics:
\begin{enumerate}
\item Each profile defines an ordered set,
\begin{align}
&\left\{\epsilon_{r_0}(\vv{x}_i,z^i_j)\middle| j=1,2,\cdots,J_i\right\},
\end{align}
  which exhibits an extremely irregular evolution that
  impedes the recognition of a continuous curve along the depth direction 
  (Fig.\,\ref{depth-ave_a}).
\item After taking the logarithm of the values, the sequences appear to be more continuous (Fig.\,\ref{depth-ave_b}).
\item
If we normalise each value with the arithmetic mean along the profile to which it belongs as follows:

\begin{align}
\varepsilon_{r_0}(\vv{x}_i,z^i_j)=
\frac{\epsilon_{r_0}(\vv{x}_i,z^i_j)}{\epsilon_L(\vv{x}_i)},~
\epsilon_L(\vv{x}_i) \define J_i^{-1}\sum_{j=1}^{J_i}\epsilon_{r_0}(\vv{x}_i,z^i_j).
\label{normalization}
\end{align}
then the histogram of the logarithmic values,
\begin{align}
 \left\{
\log{\left(
\varepsilon_{r_0}(\vv{x}_i,z^i_j)
\right)}
\middle| i=1,2,\cdots,I; j=1,2,\cdots,J_i
\right\},
\end{align}
appears as an asymmetric distribution, as we will see later in the Results section.
Note the distinction between the two symbols;
$\epsilon_{r_0}(\vv{x}_i,z^i_j)$ for the original energy dissipation rates,
and
$\varepsilon_{r_0}(\vv{x}_i,z^i_j)$ for the normalised ones.
\end{enumerate}

\subsection*{Multifractal analysis\label{scaling_analysis}}
We conduct a scaling analysis 
of the moments to derive the moment scaling exponent 
within the universal multifractal framework.
Although the analysis could be extended to 
multidimensional objects \cite{lovejoy2013weather},
the limited number of samples ($409$ profiles) prevents 
us from conducting an extensive analysis in a multidimensional framework.
Therefore,
we treat each profile as an independent sample
and analyse the statistical structure of the $1$-dimensional 
object.
\paragraph{Universal multifractal model}
The basic formulation of the universal 
multifractal model is as follows
\cite{lovejoy2013weather,gires2013development}:
Suppose we have a multifractal field, $\varepsilon_{\lambda}$, at 
resolution $\lambda$ ($=L/r$), where $r$ is the observational
scale and $L$ is the outer scale.
The field is normalised by the mean, 
that is $\left<\varepsilon_{\lambda}\right>=\left<\varepsilon\right>=1$, which
is conserved at all scales.

The probability of exceeding a scale-dependent threshold, $\lambda^{\gamma}$, 
varies according to singularity $\gamma$ as
\begin{align}
\Pr(\varepsilon_{\lambda}\geq \lambda^{\gamma})&\approx \lambda^{-c(\gamma)},
\label{extreme}
\end{align}
where $c(\gamma)$ represents the codimension function and $\approx$ represents equality up to multiplication by a slowly varying function of $\gamma$.
Thus, the multifractal model is characterised by the property that 
the codimension varies with the singularity.
This relation is 
equivalently represented as the scaling of the statistical moment of any order, $q$,
\begin{align}
\left<(\varepsilon_{\lambda})^q\right> &= \lambda^{K(q)},
\label{moment}
\end{align}
where $K(q)$ is the moment scaling function.
The two functions,
$K(q)$ and $c(\gamma)$, are actually related by the Legendre transformation
because the moment generation function 
can be written in terms of 
the occurrence probability of singular events 
using the saddle-point approximation,
$\left<\left(\varepsilon_{\lambda}\right)^q\right>
=\int \lambda^{q\gamma}\d p(\gamma)\approx \lambda^{\max_{\gamma}\left\{
q\gamma-c(\gamma)\right\}}$ \cite{ParishandFrish1985},
where $\d p(\gamma)\define 
\Pr(\lambda^{\gamma} \leq \varepsilon_{\lambda}< \lambda^{\gamma+\d\gamma})$. 
Functions $K(q)$ and $c(\gamma)$ 
determine the variability of the
multifractal field $\varepsilon_{\lambda}$ 
across the scales, $\lambda$.

Owing to a generalisation of the central limit theorem,
several multiplicative processes comprising different generators 
converge to a universal multifractal
\cite{SchertzerLovejoy1987,SchertzerLovejoy1997}, the moment exponent of which
is expressed as follows:
\begin{align}
K(q)&=\frac{C_1}{\alpha-1}(q^{\alpha}-q),\label{form_Kq}
\end{align}
where $0\leq \alpha \leq 2$ is the multifractal index and $C_1$ is the codimension of the mean.
Note that the case $\alpha = 2$ corresponds to the log-normal model advocated by the Russian school (Kolmogorov, Obukhov, Yaglom, etc.).
This equation satisfies probability normalisation, $K(0)=0$,  
and energy conservation, $K(1)=0$.
Its Legendre transformation gives:
\begin{align}
c(\gamma)&=
C_1\left(\frac{\gamma}{C_1 \alpha'}+\frac{1}{\alpha}\right)^{\alpha'},
\label{form_cg}
\end{align}
where $1/\alpha+1/\alpha'=1$.

\subsection*{Estimations based on the cascade model\label{stat_analysis}}
In this section, we discuss the 
estimation of the energy input rate, $\overline{\epsilon}$,
by utilising the information obtained from an observational profile.
While the sample mean of the energy dissipation rate along 
a profile is simply indicated 
by the arithmetic mean of the vertical data values, 
the information 
on the energy input rate and its uncertainty still needs to be obtained.
Therefore,
we will estimate the posterior distribution of 
the energy input rate from observations.
In particular, we focus on the median and confidence interval (CI).
Although the arithmetic mean over a profile is the primary 
measure for the sample, 
the characteristics of the population 
can also be evaluated by using the joint probability density 
of several different sample statistics, 
obtained from the Monte Carlo simulation of the cascade model.
The notation used in this section is summarised in Table\,\ref{notation}.
\begin{table}[ht]
\caption{Notation for the estimation study\label{notation}}
\begin{tabular}{lll}
\hline
Name & Notation & Definition\\
\hline
Index for vertical position &$j$ &$ 1,2,3,\cdots, 2^n$\\
Energy dissipation rate &$\epsilon_j$ &\\
Logarithm of energy dissipation rate &$\gamma_j$ & $\log{\epsilon_j}$\\
Energy input rate (or population mean)&$\overline{\epsilon}$ &\\
Logarithm of energy input rate &$\overline{\gamma}$ & $\log{\overline{\epsilon}}$\\
Median of estimated $\overline{\gamma}$ &$\gamma_{(0.5)}$ & 
$\Pr(\overline{\gamma}<\gamma_{(0.5)})=0.5$\\
Stable L\'{e}vy generators
& $\Gamma_{ik}$ & 
$\sim S_{\alpha}(\sigma h^{1/\alpha},-1,-\widehat{\sigma_{\alpha}}^{\alpha}h)$\\ 
Width of L\'{e}vy generator&- & $\sigma h^{1/\alpha}$ \\
Shift of L\'{e}vy generator&- & $-\widehat{\sigma_{\alpha}}^{\alpha}h$ =
$-\frac{\sigma^{\alpha}}{\cos{\left(\frac{\pi}{2}(2-\alpha)\right)}} h
=-\frac{C_1}{\alpha-1} h$ \\
Logarithm of arithmetic mean & $\widehat{\gamma}$ &
$\log{\left(2^{-n}\sum_{j=1}^{2^n}\mathrm{e}^{\gamma_j}\right)}$\\
Logarithm of geometric mean & $\widetilde{\gamma}$ &
$2^{-n}\sum_{j=1}^{2^n}\gamma_j$\\
Logarithm of quadratic mean & $\gamma^{\sharp}$ &
$2^{-1}\log{\left(2^{-n}\sum_{j=1}^{2^n} \mathrm{e}^{2\gamma_j}\right)}$\\
Marginal probability density function &$q_1$& Probability density of $\overline{\gamma}-\widehat{\gamma}$;\\
& & $q_1(\cdot)=\int \int q_3(\cdot,u,v)\d u \d v $
\\
Joint probability density function &$q_3$& Probability density of 
$(\overline{\gamma}-\widehat{\gamma},
\widehat{\gamma}-\widetilde{\gamma},
\widehat{\gamma}-\gamma^{\sharp})$
\\
\end{tabular}
\end{table}

\begin{table}[ht]
\caption{Constants and parameters for the estimation study\label{params}}
\begin{tabular}{llr}
\hline
Meaning & Parameter   & Value\\
\hline
Number of steps in cascade & $n$ & $8$\\
Number of vertical points & $2^n$ &$256$\\
Step size of cascade &$h$ & $\log{2}$\\
Number of samples in Monte Carlo simulation & $M$ &$1.024\times 10^{10}$\\
Multifractal index & $\alpha$ & $1.62$\\
Codimension of the mean & $C_1$ & $0.352$\\
Width of bins in histogram of $q_3$ & $(w_1,w_2,w_3)$ 
& $(0.1, 0.1, 0.05)$ \\
Number of bins in histogram of $q_3$ & $(n_1,n_2,n_3)$ 
&$(500,200,200)$ \\
Number of profiles in identical twin exp.&$M'$ &$30000$\\
Number of observed profiles (Obs.)&I &$409$\\
Obs. with more than $2^n$ vertical points &- &$353$\\
\end{tabular}
\end{table}

\begin{figure}[ht]
 \begin{center}
  \includegraphics[width=0.9\columnwidth]{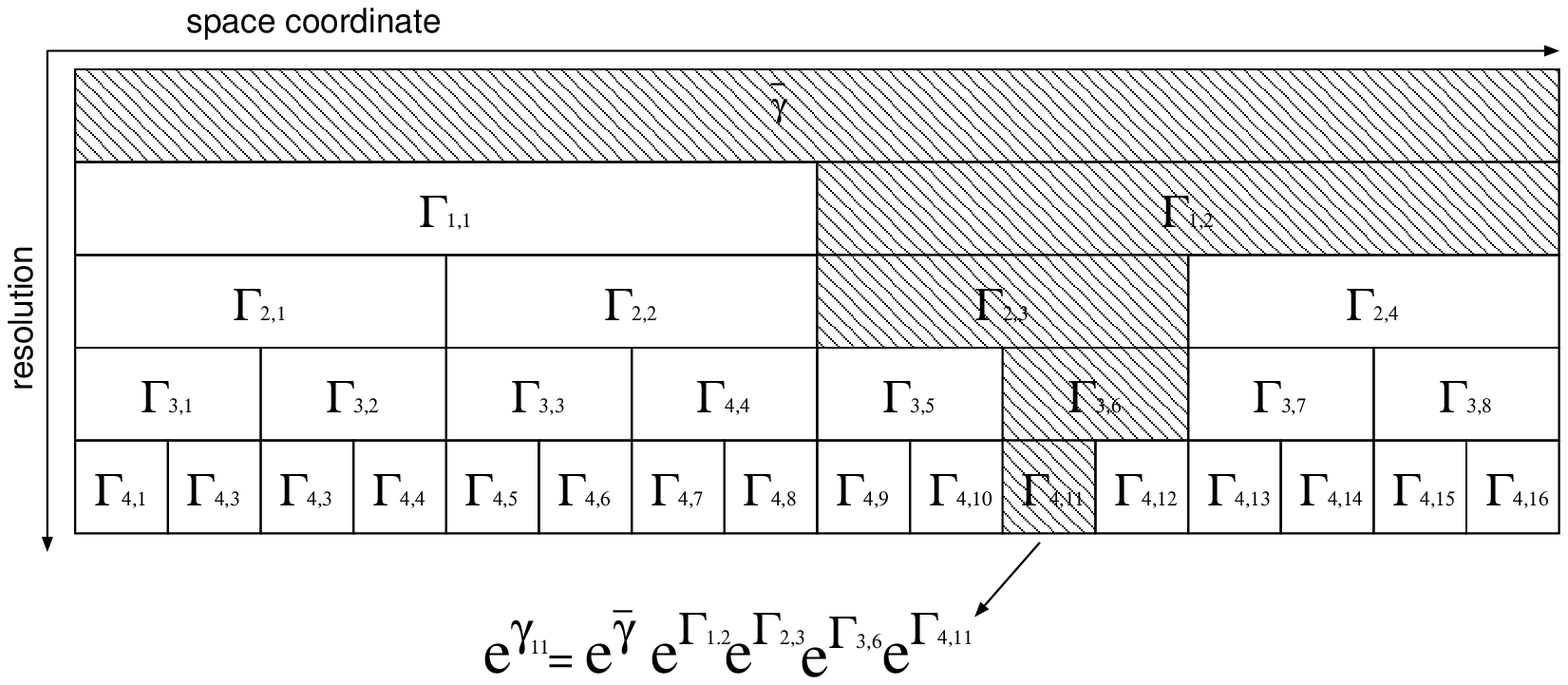}
  \caption{
  Schematic of the multiplicative cascade model.
  The energy dissipation rate at $z_{11}$ at resolution $r_4=L/2^4$ is considered as an example.
  \label{cascade}}
 \end{center}
\end{figure}

\paragraph{Multiplicative cascade simulation\label{sec_cascade}}
To examine the relationship between various statistical quantities 
derived from observational profiles,
we construct a simulation model for the multiplicative cascade 
by following the procedure described in Schmitt (2003)
\cite{schmitt2003modeling}, 
as shown in Fig.\,\ref{cascade}.
Each building block, $\Gamma_{ik}$, is a generator that obeys a left-skewed
stable distribution, $S_{\alpha}(\sigma h^{1/\alpha},-1,-\widehat{\sigma_{\alpha}}^{\alpha}h)$,
with $h=\log{2},~
\widehat{\sigma_{\alpha}}^{\alpha}\define
\sigma^{\alpha} /\cos{\left(\frac{\pi}{2}(2-\alpha)\right)}
=C_1/(\alpha-1)$ \cite{samorodnitsky1994non}. 

Consider a fixed horizontal position $\vec{x}$.
Let 
$\overline{\epsilon}=\exp{\left(\overline{\gamma}\right)}$ 
be the energy input rate for a profile at $\vec{x}$,
$n=\log_{2}\frac{L}{r}$ be the number of steps,
$0 \leq n' \leq n$ be the scale index, and
$1 \leq j \leq 2^n$ be the spatial index. 
The cascade simulation is performed for variable $X_{n',j}$ as follows.
\begin{enumerate}
 \item For each spatial index $j=1,2,\cdots,2^n$, set $X_{0,j}=\overline{\gamma}$.
 \item For each scale index $n'=1,\cdots,n$, repeat the following steps: 
       \begin{itemize}
	\item For each spatial block $k=1,2,\cdots,2^{n'}$, perform the following steps:
	\begin{enumerate}
	 \item Generate a random variable, $\xi_{n'k}$, which obeys $S_{\alpha}(1,-1,0)$ \cite{Misiorek2012}.
	 \item For each spatial index $j=(k-1)\cdot 2^{n-n'}+1,\cdots,k\cdot 2^{n-n'}$, downscale $X$ by
\begin{align}
X_{n',j}&=
X_{n'-1,j}+\Gamma_{n',k},\quad
\Gamma_{n',k}\define
-\widehat{\sigma_{\alpha}}^{\alpha} h + \sigma h^{\frac{1}{\alpha}} \xi_{n'k}.\label{defineX}
\end{align}	       
	\end{enumerate}
\item 
For each spatial index $j=1,2,\cdots,2^n$, set $\gamma_j=X_{n,j}$.
       \end{itemize}
\end{enumerate}

The output, $\gamma_j$, represents
the logarithm of the energy dissipation rate at the horizontal position, 
$\vec{x}$, and the vertical position, 
$z_j \in [(j-1) r_n, j r_n]$, at 
the resolution $r_n=L/2^n$. 
By using the floor function,
the cascade process 
can be more compactly represented as
\begin{align}
\gamma_j&=
\overline{\gamma}
+\sum_{n'=1}^{n} \Gamma_{n',\lfloor (j-1)/2^{n-n'} \rfloor+1},
\quad j=1,2,\cdots,2^n.\label{cas}
\end{align}

An important implication of this formulation 
is that 
the arithmetic mean of the vertical data points
is not necessarily equal to the 
the energy input rate
because the cascade process has a fluctuating nature.
In other words, a realisation of the 
vertical average, $\epsilon_L
=2^{-n}\sum_{j=1}^{2^n} \exp{\left(\gamma_j\right)}$, 
is not always equal to $\exp{(\overline{\gamma}})$,
whereas the expectation $\E\left[\epsilon_L\right]$ is; 
hence, we can regard the latter as the population mean for a profile.
Below, we focus mainly on the relationship between 
the arithmetic mean over a profile
and the energy input rate. 
We perform statistical estimations 
from one to the other of these quantities 
based on the cascade model.

\paragraph{Estimation of the energy input rate}\label{est_inp}
In this subsection, we first describe the statistical relationship between 
the population mean and various moments over a profile,
based on the cascade model.
Then, we derive a formula for 
the posterior probability given the observation of the moments.
Finally, we use this formula as the basis of a concrete procedure
for computing the posterior probability. 

We assume that each set of $\gamma_j$'s is generated by an $n$-step cascade model 
as in Eq.\,(\ref{cas}).
Here, we want to estimate 
the energy input rate $\overline{\gamma}$,
which corresponds to the population mean, 
by using the information from 
the observed data $\{\gamma_j|j=1,2,\cdots,2^n\}$.
In this regard, 
in addition to the arithmetic mean $\widehat{\gamma}$,
which corresponds to $K(1)$ in Fig.\,\ref{multi},
we can also use other moments over a profile, e.g., the geometric mean $\widetilde{\gamma}$ and 
quadratic mean $\gamma^{\sharp}$, which correspond to $K'(0)$ and $K(2)$, respectively.
We can derive the following expressions based on Eq.\,(\ref{cas}):
\begin{align}
\widehat{\gamma}&=\overline{\gamma} +\log{\left\{
2^{-n}\sum_{j=1}^{2^n}\exp{\left(\sum_{n'=1}^{n}\Gamma_{n',
\lfloor (j-1)/2^{n-n'} \rfloor+1}\right)}\right\}},\\
\widetilde{\gamma}&=\overline{\gamma} +
2^{-n}\sum_{j=1}^{2^n}\sum_{n'=1}^{n}\Gamma_{n',\lfloor (j-1)/2^{n-n'} \rfloor+1},\\
\gamma^{\sharp}&=\overline{\gamma} +\frac12\log{\left\{
2^{-n}\sum_{j=1}^{2^n}\exp{\left(2\sum_{n'=1}^{n}\Gamma_{n',\lfloor (j-1)/2^{n-n'}
\rfloor+1}\right)}\right\}},
\end{align}
where we find that the term $\overline{\gamma}$ is factored out.
Therefore, $\overline{\gamma}-\widehat{\gamma}$,
$\widehat{\gamma}-\widetilde{\gamma}$, and 
$\widehat{\gamma}-\gamma^{\sharp}$
are independent of $\overline{\gamma}$,
and thus dimensionless.

The structure of the cascade model implies that 
the appearance probability of $\widehat{\gamma}$ given $\overline{\gamma}$
is determined only by their difference:
$P(\widehat{\gamma}|\overline{\gamma})
=q_1(\overline{\gamma}-\widehat{\gamma})$.
Furthermore, by assuming that we have no prior information 
about $\overline{\gamma}$, Bayes' theorem is applied
to invert it into the posterior probability for $\overline{\gamma}$ as follows.
\begin{align}
P(\overline{\gamma}|\widehat{\gamma})&=
\frac{P(\widehat{\gamma}|\overline{\gamma})P(\overline{\gamma})}
{\int
P(\widehat{\gamma}|\overline{\gamma})P(\overline{\gamma})
\d\overline{\gamma}}
=
q_1(\overline{\gamma}-\widehat{\gamma}).
\label{est1i}
\end{align}

Furthermore, we can extract information from  
$\widetilde{\gamma}$ and $\gamma^{\sharp}$.
They are encoded in the joint probability density function (PDF)
$q_3(\overline{\gamma}-\widehat{\gamma},\widehat{\gamma}-\widetilde{\gamma},
\widehat{\gamma}-\gamma^{\sharp})$ 
computed from Monte Carlo simulation of the cascade model.
Then, a conditional PDF is derived as
\begin{align}
q_3(\overline{\gamma}-\widehat{\gamma}|\widehat{\gamma}-\widetilde{\gamma},
\widehat{\gamma}-\gamma^{\sharp})
&=
\frac{
q_3(\overline{\gamma}-\widehat{\gamma},
\widehat{\gamma}-\widetilde{\gamma},
\widehat{\gamma}-\gamma^{\sharp}
)
}{
\int q_3(\overline{\gamma}-\widehat{\gamma},\widehat{\gamma}-\widetilde{\gamma},
\widehat{\gamma}-\gamma^{\sharp}) \d\overline{\gamma}
}.\label{cond3d}
\end{align}
A procedure similar to Eq.\,(\ref{est1i})
can be applied to obtain another posterior probability:
\begin{align}
P(\overline{\gamma}|\widehat{\gamma},
\widehat{\gamma}-\widetilde{\gamma}=u,
\widehat{\gamma}-\gamma^{\sharp}=v)
&=
q_3(\overline{\gamma}-\widehat{\gamma}|u,v),
\label{est1c}
\end{align}
under the constraints 
$\widehat{\gamma}-\widetilde{\gamma}=u,
\widehat{\gamma}-\gamma^{\sharp}=v$.

On the basis of the above formulation, 
we perform an identical twin experiment obeying the following procedure.
\begin{enumerate}
\item Perform a Monte Carlo experiment to obtain 
$q_3(\overline{\gamma}-\widehat{\gamma},
 \widehat{\gamma}-\widetilde{\gamma},
 \widehat{\gamma}-\gamma^{\sharp})$.
\begin{enumerate}
 \item 
Set the energy input rate to $\overline{\gamma}=0$.
\item \label{gen_samples}
Create many random samples of the profile 
using the cascade model in Eq.\,(\ref{cas}).
\item
Add up the frequency of occurrence to 
derive the joint PDF
$q_3(\overline{\gamma}-\widehat{\gamma},
 \widehat{\gamma}-\widetilde{\gamma},
 \widehat{\gamma}-\gamma^{\sharp})$.
\end{enumerate}
 \item \label{item1}
Set the energy input rate $\overline{\gamma}$ to a random number.
\item \label{item2}
Create a random pseudo-observation sample of the profile 
using the cascade model in Eq.\,(\ref{cas}).
\item
Calculate the statistics $\widehat{\gamma},
 \widehat{\gamma}-\widetilde{\gamma},
 \widehat{\gamma}-\gamma^{\sharp}$ for the profile.
\item
Compute the conditional PDF
$P(\overline{\gamma}|\widehat{\gamma},
\widehat{\gamma}-\widetilde{\gamma},\widehat{\gamma}-\gamma^{\sharp})$.
 \item 
Calculate the median and $95\%$ CI for 
the estimated $\overline{\gamma}$.
 \item 
Compare the estimate of $\overline{\gamma}$
with its true value.
\end{enumerate} 
The same procedure is applied to the real data experiment, 
except that $\overline{\gamma}$ in \ref{item1} is unknown, 
as follows.
\begin{enumerate}
\item Pick an observed profile, 
and 
calculate the statistics $\widehat{\gamma},
 \widehat{\gamma}-\widetilde{\gamma},
 \widehat{\gamma}-\gamma^{\sharp}$ for the profile.
\item
Compute the conditional PDF
$P(\overline{\gamma}|\widehat{\gamma},
\widehat{\gamma}-\widetilde{\gamma},\widehat{\gamma}-\gamma^{\sharp})$,
using the joint PDF
$q_3(\overline{\gamma}-\widehat{\gamma},
 \widehat{\gamma}-\widetilde{\gamma},
 \widehat{\gamma}-\gamma^{\sharp}
)$ obtained from the Monte Carlo experiment.
\item 
Calculate the median and $95\%$ CI for 
the estimated $\overline{\gamma}$.
\end{enumerate} 
Among the indices for the estimated result,
the conditional expectation of 
$\overline{\epsilon}=\exp{(\overline{\gamma})}$ 
is not necessarily defined as a finite value
because the posterior distribution of $\overline{\gamma}$ 
is neither Gaussian nor left-skewed stable.
In contrast, the percentiles, 
including the median and the $95\%$ CI, 
are always defined for the distribution. 
They are also preserved,
$\Pr(\overline{\gamma}<a)=\Pr(f(\overline{\gamma})<f(a))$,
 under the 
increasing transformation $f:\overline{\gamma}
 \mapsto \exp{(\overline{\gamma})}$.
We therefore employ the median and the $95\%$ CI 
as robust indices.

To confirm that the uncertainty in the estimated values 
of $\alpha$ and $C_1$ 
does not diminish the performance of the proposed method,
we treat these parameters as random variables
(specified in the results), 
when making pseudo-observation 
samples in the identical twin experiments (procedure\,\ref{item2}).

We have thus 
established a procedure for estimating the population mean for a profile 
based on the joint probability distribution of several moments over a profile 
computed by a Monte Carlo simulation of the cascade model.

\section*{Results}
\subsection*{Analysis of observational data \label{ana}}
Suppose we have 
the observational data 
of the normalised energy dissipation rate, 
$\varepsilon_{r_0}(\vv{x})$,  
in bin width $r_0$ at the horizontal position $\vv{x}$,
as well as their spatial average $\varepsilon_{r}(\vv{x})$ in width $r \geq r_0$.
In terms of the universal multifractal model
(\ref{moment}),
the scaling of the statistical moments
in the data takes the form 
\begin{align}
\frac{\left<\varepsilon_{r_0}(\vv{x})^q \right>}
     {\left<\varepsilon_{r} (\vv{x})^q \right>}
&= \left(\frac{r}{r_0}\right)^{K(q)},
 \label{q_Kq}
\end{align}
where $\left<\cdot\right>$ denotes the expected value.
This implies that 
the expectation of the $q$-th moment at a scale 
over the one at another scale should be equal 
to the $K(q)$-th power of the resolution ratio,
regardless of the horizontal position $\vv{x}$.
We approximate the expected value in Eq.\,(\ref{q_Kq}) with the empirical average
\begin{align} 
\left<\varepsilon_{r}(\vv{x})^q\right> 
&\fallingdotseq
\frac{\sum_{i=1}^I\sum_{k=1}^{J_i(r)}
\left(\varepsilon_{r}^{(i,k)}\right)^q}{\sum_{i=1}^I J_i(r)},\quad
\varepsilon_{r}^{(i,k)}
\define
2^{-n}\sum_{j=2^n(k-1)+1}^{2^n k}\varepsilon_{r_0}(\vv{x}_i,z^i_j),
\label{slope}
\end{align}
where 
$r=2^n r_0$ is a resolution larger than or equal to $r_0$, and
$\varepsilon_{r_0}(\vv{x}_i,z^i_j)$ is 
the normalised value defined in Eq.\,(\ref{normalization}).
The superscript $(i,k)$ runs across 
all profiles indexed by $i$, each of which 
has total $J_i(r)$ segments in resolution $r$.
By substituting Eq.\,(\ref{slope}) into Eq.\,(\ref{q_Kq}),
we can evaluate the values of $K(q)$ according to $q$.

\begin{figure}[ht]
  \begin{center}
    \includegraphics[width=0.75\columnwidth]{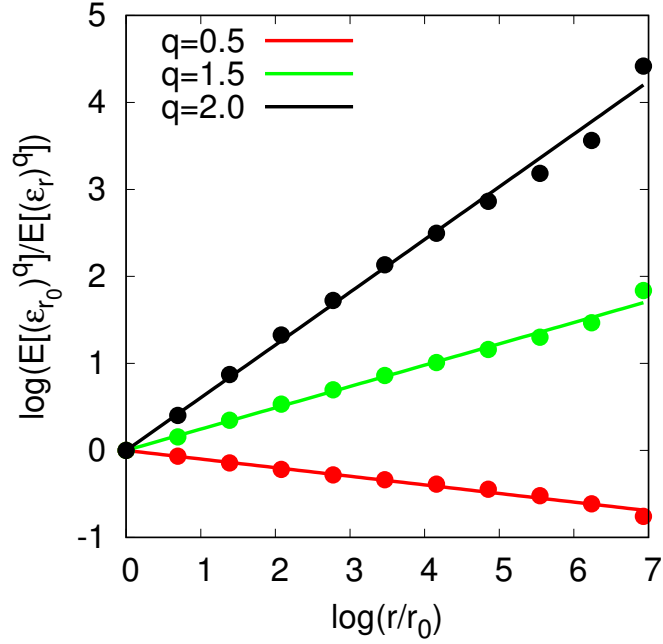}
\caption{Scale dependency of the moments, 
$\left(\log{(r/r_0)},-\log{\left<(\varepsilon_r/\varepsilon_{r_0})^{q}\right>}\right),$
where $r_0$ is the width of the observational bin.
The moment scaling exponents are found to be 
$K(0.5)=-0.099\pm0.003,~K(1.5)=0.245\pm0.007,~K(2.0)=0.606\pm0.017.$
\label{scale_dep}}
  \end{center}
\end{figure}

\begin{figure}[ht]
  \begin{center}
    \includegraphics[width=0.75\columnwidth]{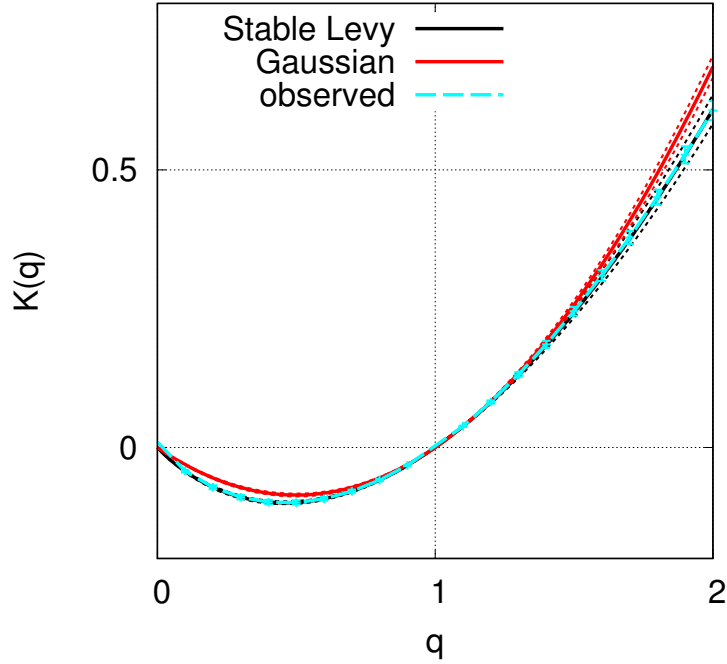}
\caption{Moment scaling exponent $K(q)$ 
for observational data (cyan).
Best-fitting multifractal model with stable L\'{e}vy generators (black),
and 
with Gaussian generators (red).
Each error bar in cyan shows the standard deviation 
for the fitting of $K(q)$.
Dotted lines in black and red indicate the ranges of error 
due to the uncertainty of parameters in the corresponding models.
\label{multi}}
  \end{center}
\end{figure}

The scalings for several moments are shown in Fig.\,\ref{scale_dep}.
Using the various slope values,
the observational curve of $(q,K(q))$ 
in the range of $0 \leq q\leq 2$ 
is indicated in Fig.\,\ref{multi} in cyan.
We can estimate the parameters, $\alpha$ and $C_1$,
by fitting the theoretical curve\,(\ref{form_Kq}) to the observational curve.
To consider the uncertainty 
in the observational curve,
we used the bootstrap method \cite{efron1994introduction}
with $1000$ trials, each of which 
has $409$ profiles that are randomly sampled with replacements
from the original set of $409$ profiles.
We thereby obtained the parameters
$\alpha = 1.62 \pm 0.03,\quad
C_1 = 0.352 \pm 0.009$  
for the multifractal model with stable L\'{e}vy generators, 
i.e., the universal multifractal model.
By taking into account the dependency on $\alpha$, 
we can also estimate $C_1$ as $C_1=0.109\alpha+0.175\pm 0.008$.
By a similar procedure, we obtain 
the parameter $C_1=0.343\pm 0.010$ 
for the multifractal model with Gaussian generators,
corresponding to the original Yaglom cascade with $\mu=2C_1$.
In Fig.\,\ref{multi}, the observational curve (cyan)
and the theoretical curve 
for the multifractal model with stable L\'{e}vy generators (black)
are in good agreement, while the theoretical curve
for the multifractal model with Gaussian generators, i.e., the log-normal model (red),
has a different curvature from the observational curve.

The parameter values for the multifractal model with stable L\'{e}vy generators
are largely 
consistent with previous results for atmospheric dissipation 
fields 
($\alpha=1.35\pm 0.07,~C_1=0.3\pm 0.05$ for the horizontal shear of a velocity field \cite{npg-1-105-1994};
$\alpha=1.85\pm 0.05,~C_1=0.59\pm 0.05$ for vertical kinetic energy flux \cite{npg-1-115-1994}).

\begin{figure}[ht]
  \begin{center}
    \includegraphics[width=0.75\columnwidth]{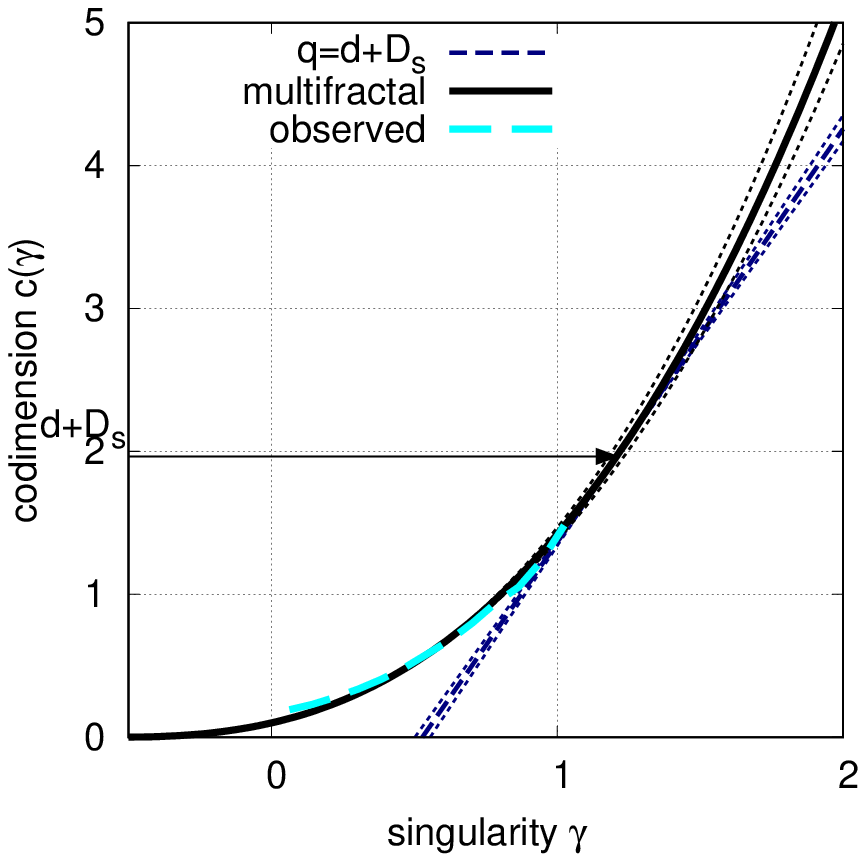}
\caption{Codimension $c(\gamma)$ of singularities $\gamma$
for the best-fitting multifractal model with stable L\'{e}vy generators (black). 
The corresponding curve for the observational data 
is shown for reference (cyan).
Sampling dimension $D_s$ and 
the limitation for the moment exponent 
(the slope of the navy-blue line) are also shown. 
Dotted lines in black and navy indicate the ranges of error 
due to the uncertainty in the model parameters.
\label{rare}}
  \end{center}
\end{figure}

Figure \,\ref{rare} shows the theoretical curve of extremes for
the multifractal model (\ref{form_cg})
in black
and the observational curve,
\begin{align}
c_{\text{obs}}(\gamma)&=-\log_{\lambda}\left[
g(\gamma){\Pr\left(\varepsilon_{r_0}>
\lambda^{\gamma}\right)} \right],\quad \lambda=L/r_0,\label{cobs}
\end{align}
in cyan, where $\lambda=2^9$ is used;
this is a typical scale ratio in the data.
Note that the correction term,
\begin{align}
g(\gamma)&=\sqrt{2\pi\alpha c(\gamma) \log{\lambda}},\label{prefactor}
\end{align}
compensates for the prefactor in the asymptotic 
complementary cumulative distribution function, 
$g(\gamma)^{-1}\mathrm{e}^{-c(\gamma)}$
\cite[Eq.\,1.2.11]{samorodnitsky1994non}.
The two curves (\ref{form_cg}) and (\ref{cobs})
 appear to be in general agreement, 
except for a slight discrepancy that is possibly 
due to the ambiguity in the selected scale ratio, $\lambda$.
Moreover, 
as our data have the sampling dimension 
\cite{lovejoy2013weather}
$D_s=\log_{\lambda}{N_s}\simeq \log{409}/\log{(2^9)}=0.963$,
the upper bound for $q$ is calculated to be $q_s=2.89\pm 0.11$ (the slope of the navy-blue 
line in Fig.\,\ref{rare}), 
which justifies the range we set ($0 \leq q\leq 2$).

\begin{figure}[ht]
  \begin{center}
    \includegraphics[width=0.7\columnwidth]{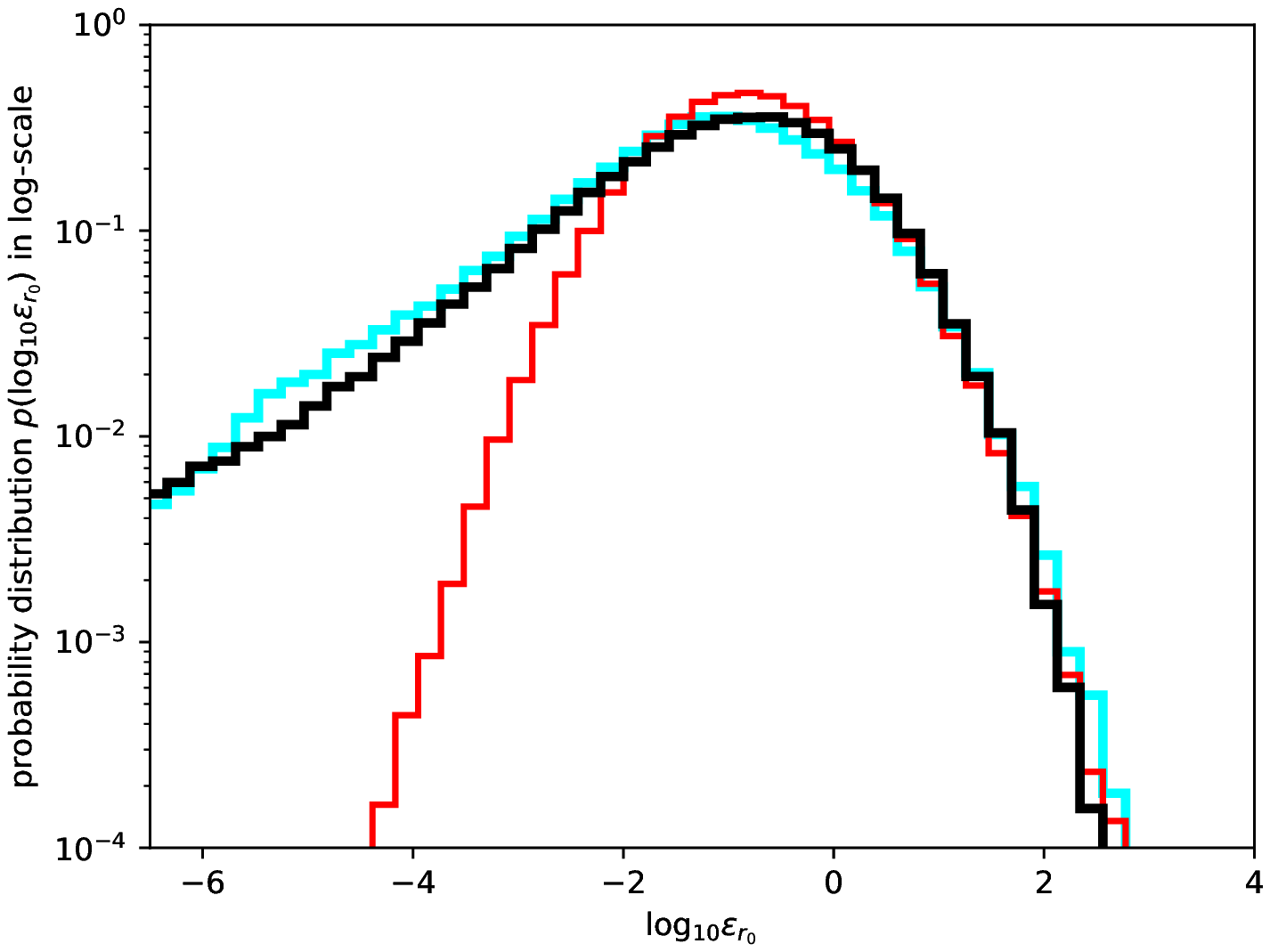}

    \caption{Distribution of the logarithm of
   observational data normalised for each profile (cyan), and 
      comparison with the statistics of samples generated from
   multiplicative cascade with 
   Gaussian/stable L\'{e}vy generators (red/black).
 \label{sing}}
  \end{center}
\end{figure}

To demonstrate the appropriateness of the 
universal multifractal model, the histogram for the logarithm of 
the bin values in the observational data 
is shown in Fig.\,\ref{sing} and compared with the samples from multiplicative 
cascade models.
Each bin value
is normalised by the arithmetic mean 
along the profile it belongs to: $\varepsilon_{r_0}=\epsilon_{r_0}/\epsilon_L.$
The histogram for the logarithm of bin data, 
$\log_{10}{\varepsilon_{r_0}}$,
appears to be in good agreement with the histogram 
of samples generated by the $8$-step cascade model 
with stable L\'{e}vy generators (black; $\alpha = 1.62, ~C_1=0.352$) 
and in poor agreement with that
generated by the multifractal model with Gaussian generators, i.e., the log-normal model 
(red; $\alpha = 2, ~C_1=0.343$). 

Moreover, in the same manner as 
the correlation in Yaglom's cascade,
the observational profiles have a power-law autocorrelation,
\begin{align}
  \left<\varepsilon(z)\varepsilon(z+\ell)\right>&\propto 
  \ell^{-K(2)}=\ell^{-0.609}, ~\ell>0,
  \label{corr_obs}
\end{align}
where $\varepsilon(z)$ is the energy dissipation rate at depth $z$.
The negative exponent explains the discontinuous characteristics 
observed in the profiles (see Fig.\,\ref{depth-ave}a).

\subsection*{Simulations of cascade model}
Before estimating 
the energy input rate $\overline{\epsilon}$
that corresponds to each observational profile,
we first performed a Monte Carlo experiment with simulations
of $1.024\times 10^{10}$ particles (profiles)
using the $8$-step cascade model.
The constants and parameters used 
in the simulation and estimation study 
are summarised in Table\,\ref{params}.

\begin{figure}
    \begin{center}
     \includegraphics[width=1.0\columnwidth]{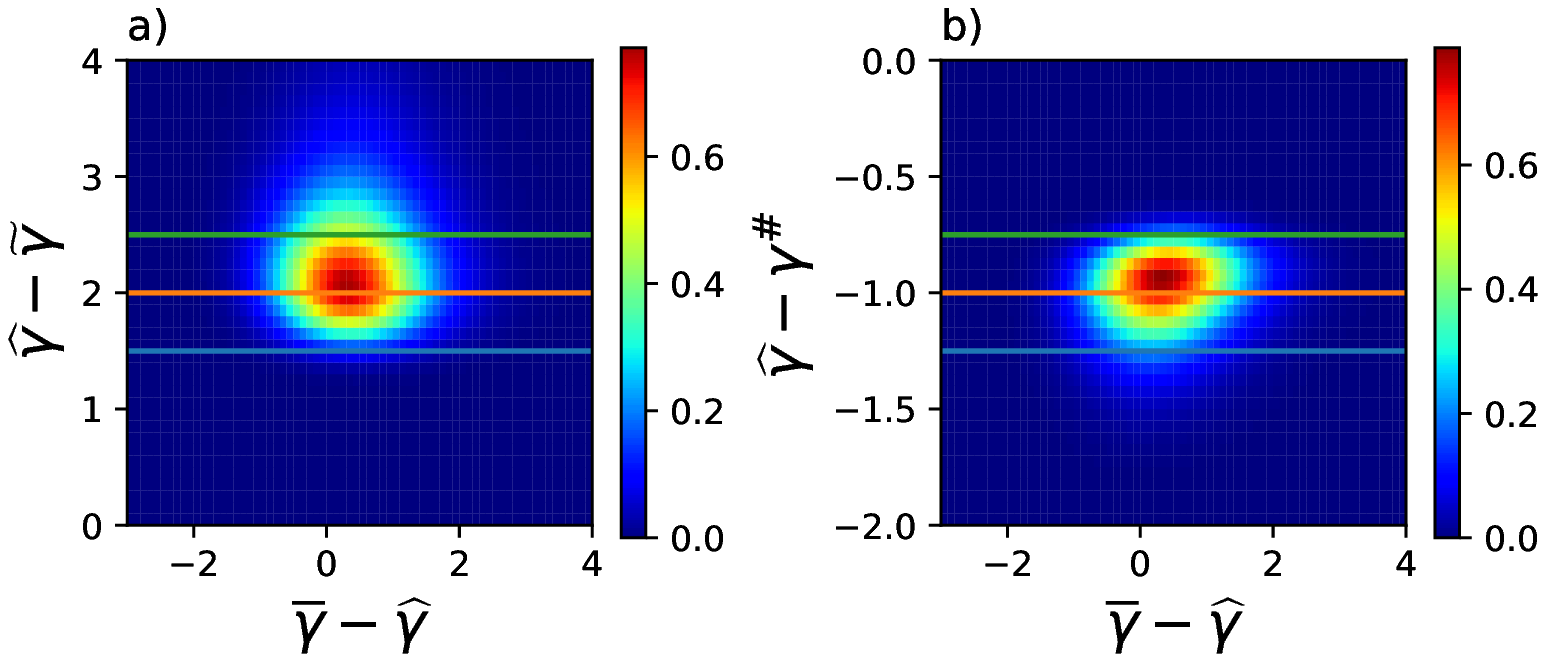}\label{histo3d}
\caption{Examples of cross section cut of joint probability density function
$q_3(\overline{\gamma}-\widehat{\gamma},
\widehat{\gamma}-\widetilde{\gamma},
\widehat{\gamma}-\gamma^{\sharp})$.
a) section cut $q_3(\cdot,\cdot,-1)$
with section lines $\widehat{\gamma}-\widetilde{\gamma}=1.5, 2, 2.5$, and 
b) section cut $q_3(\cdot,2,\cdot)$
with section lines $\widehat{\gamma}-\gamma^{\sharp}=-1.25, -1, -0.75$.
\label{p3d}
}
    \end{center}
  \end{figure}

\begin{figure}
    \begin{center}
     \includegraphics[width=0.5\columnwidth]{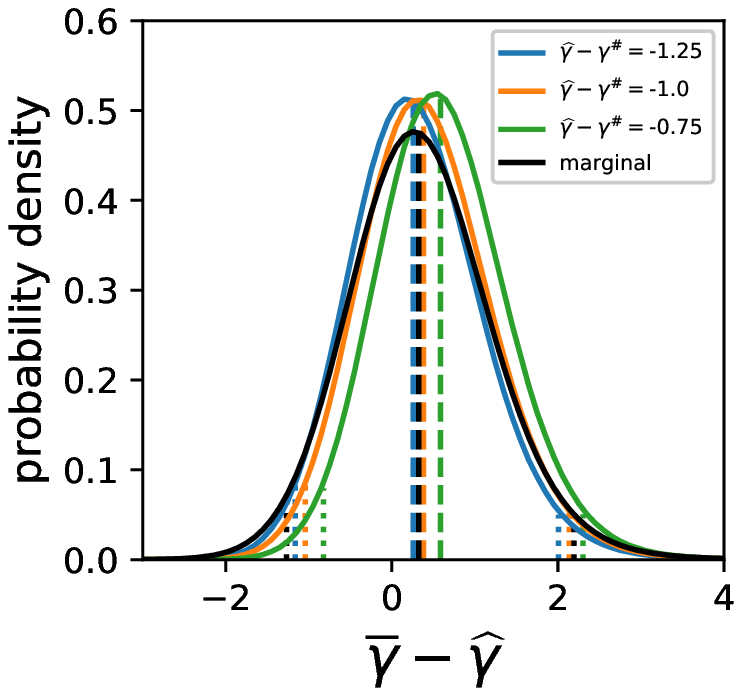}
\caption{Examples of 
conditional probability density function $q_3(\cdot|u,v)$ (coloured) and marginal probability 
distribution $q_1(\cdot)$ (black),
along with median (dashed lines) and $95\%$ confidence intervals (dotted lines).
The conditional probabilities are for $u=2; v=-1.25, -1, -0.75$.\label{perc3d}}
    \end{center}
  \end{figure}

For each particle (or profile), we generate random numbers 
$\{\gamma_j|j=1,2,\cdots,256\}$ 
from $\overline{\epsilon}=1$ according to 
the procedure discussed in Methods, and we 
add up the histograms for all the particles into 
the joint PDF
$q_3(\overline{\gamma}-\widehat{\gamma},
\widehat{\gamma}-\widetilde{\gamma},
\widehat{\gamma}-\gamma^{\sharp})$
(Fig.\,\ref{p3d}).
Examples for conditional PDF $q_3(\overline{\gamma}-\widehat{\gamma}|
\widehat{\gamma}-\widetilde{\gamma},
\widehat{\gamma}-\gamma^{\sharp})$
are shown in Fig.\,\ref{perc3d}.

\subsection*{Identical twin experiment}
Second, we estimated the posterior probability distribution for 
the energy input rate, $\overline{\epsilon}$ by inverting the probability 
distribution computed from the simulation of the cascade model.
Before applying this to real data, we performed 
an identical twin experiment using pseudo-observational data, 
whose energy input rates were given manually; thus, the estimation result 
could be checked against them.
The  inversion was performed using the result of
 the joint PDF $q_3$ created by the cascade model with the L\'{e}vy generator
(case {\tt L3}).
When generating the pseudo-observational data, 
the parameters for each profile were set randomly 
as $\alpha = 1.62 + 0.03\xi_1,~
C_1=0.109\alpha+0.175 + 0.008\xi_2,$
using standard normal random numbers, $\xi_1$ and $\xi_2$.

\begin{figure}
    \begin{flushleft}\hspace{4em} a) \hspace{26em} b) \end{flushleft}
    \begin{center}
     \includegraphics[width=0.49\columnwidth]{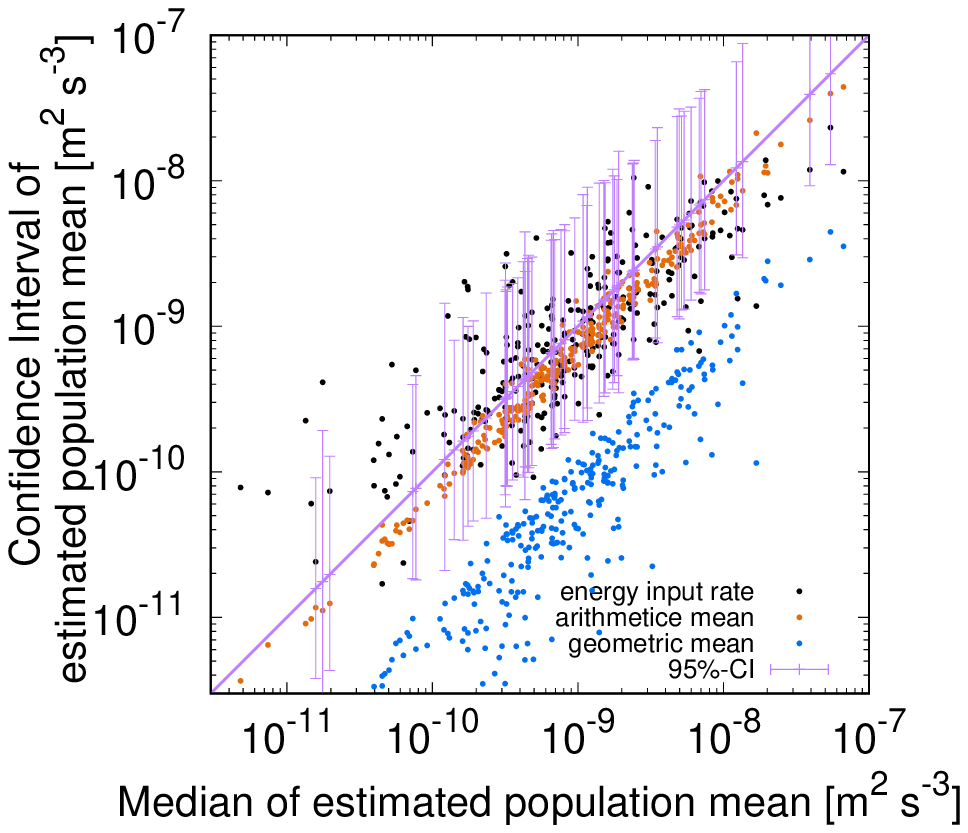}
     \includegraphics[width=0.49\columnwidth]{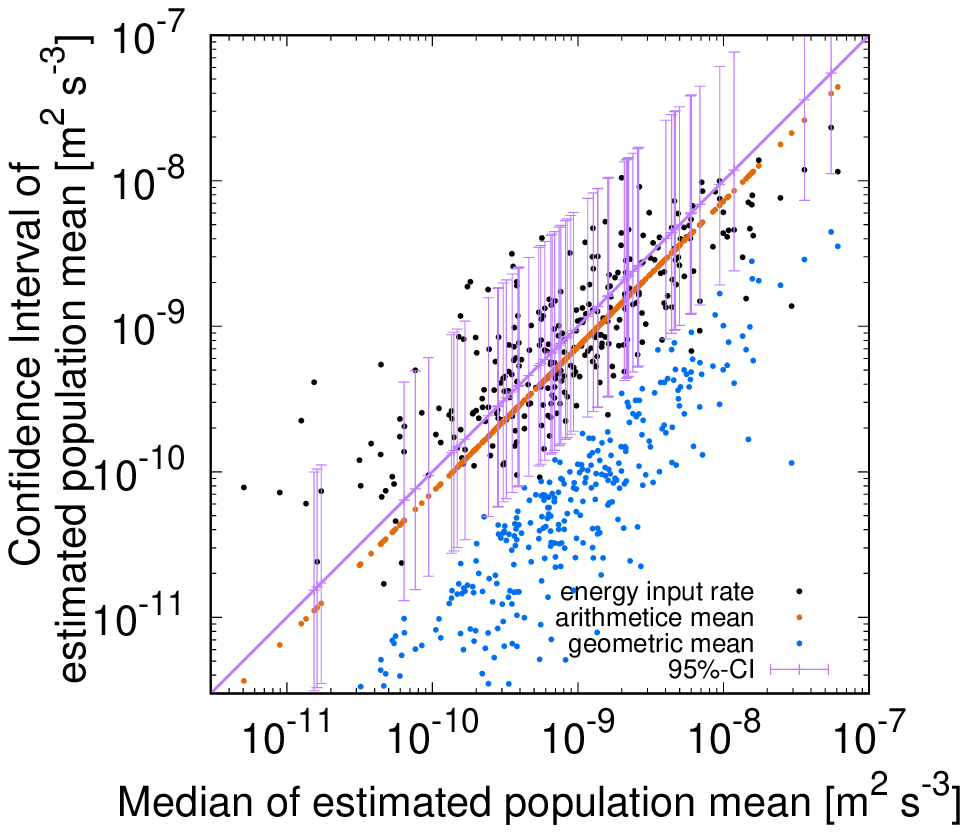}
     \caption{Result of identical twin experiments.
     Median (horizontal axis) versus confidence interval (vertical axis) 
     of $\overline{\epsilon}$.
     a) Result using probability density function $q_3(\cdot|u,v)$.
     b) Result using probability density function $q_1(\cdot)$.
     For given values of the median on the horizontal axis,
     the points on the vertical axis indicate the values of 
     the confidence interval (purple segment), arithmetic mean (orange), 
     geometric mean (blue), and energy input rate (black). 
     For readability, $300$ points and $60$ intervals are
     drawn out of $30000$ trials.
     \label{est}}
    \end{center}
   \end{figure}

The result of the identical twin experiment using 
$q_3(\overline{\gamma}-\widehat{\gamma}|\widehat{\gamma}-\widetilde{\gamma},
\widehat{\gamma}-\gamma^{\sharp})$
is shown in Fig.\,\ref{est}a.
In $28497$ trials out of $30000$ (about $95\%$),
the true value of $\overline{\gamma}$ lies within the CI,
which ensures the validity of the estimation method.

\subsection*{Control experiments} 
\begin{figure}
    \begin{flushleft}\hspace{4em} a) \hspace{26em} b) \end{flushleft}
    \begin{center}
     \includegraphics[width=0.49\columnwidth]{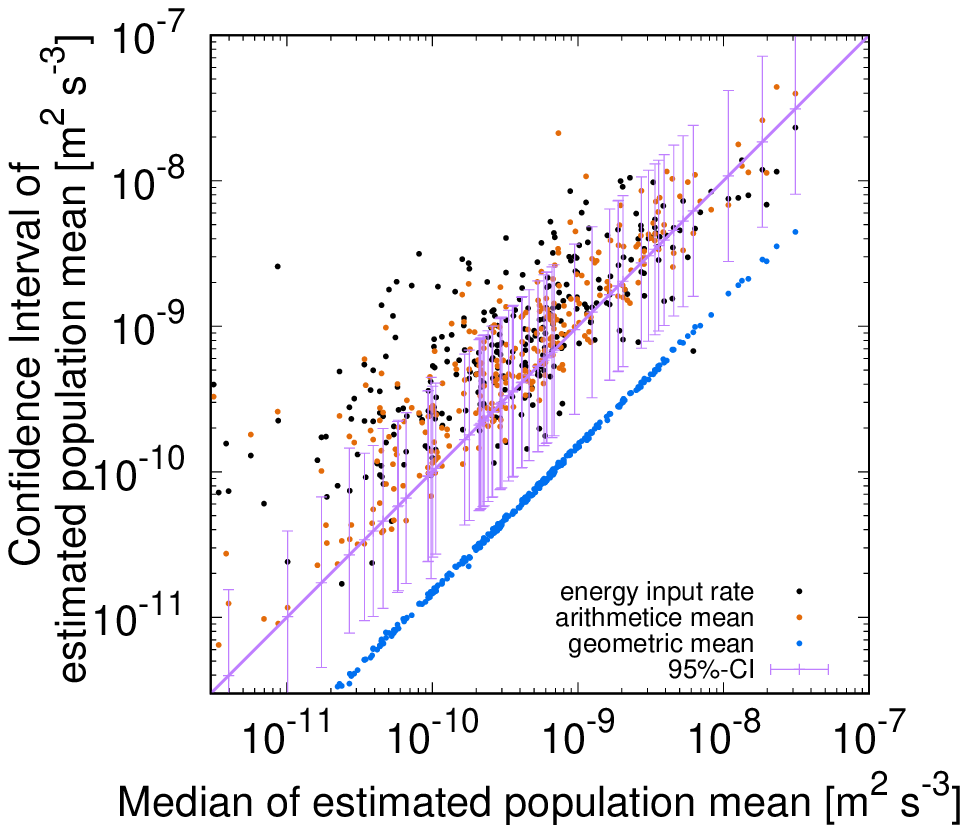}
     \includegraphics[width=0.49\columnwidth]{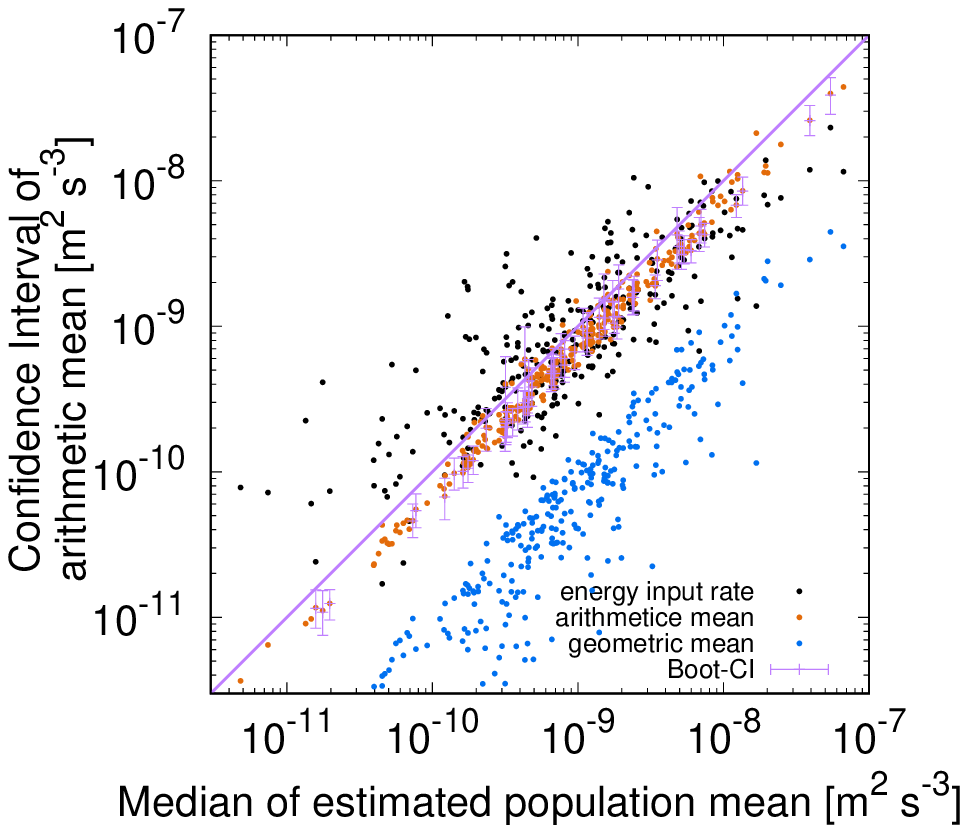}
    \caption{Result of control experiments.
     Median (horizontal axis) versus confidence interval (vertical axis)
     of $\overline{\epsilon}$,
     which are obtained 
     using (a) joint probability density function based on 
     the cascade model with Gaussian generators, and (b) the bootstrap method.
     For given values of the median on the horizontal axis,
     the points on the vertical axis indicate the values of 
     the confidence interval (purple segment), arithmetic mean (orange), 
     geometric mean (blue), and energy input rate (black). 
     For readability, $300$ points and $60$ intervals are
     drawn out of $30000$ trials.
     \label{est_gauss}}
    \end{center}
   \end{figure}

To show the superiority of the proposed method ({\tt L3}), 
which performs an inversion using the result of the joint PDF $q_3$ created by the cascade model with the L\'{e}vy generator,
the generated samples above are also estimated by other methods:
an inversion based on the marginal PDF $q_1$,
an inversion based on 
the joint PDF generated by a multiplicative cascade with 
Gaussian generators,
and a simple bootstrap method.
\paragraph{Error estimation based on 
the PDF of the arithmetic mean}
Using several different moments for a profile 
should be effective for precise error estimation.
To check this,
we perform a control experiment using 
the marginal PDF, $q_1$, of the arithmetic mean (case {\tt L1}).
We take samples created by the cascade model 
with the stable L\'{e}vy generators, 
and then estimate the CI
via the marginal PDF $q_1(\overline{\gamma}-\widehat{\gamma})$.
The results are shown in 
Fig.\,\ref{est}b, where the CI and median 
show a common positional relation to the arithmetic mean.
Comparing the case with $q_3$ to the one with $q_1$,
$75\%$ of trials in the former have narrower CIs.
This indicates that 
using information from $\widetilde{\gamma}$ and $\gamma^{\sharp}$
improves the error estimation.
 \paragraph{Error estimation based on a cascade model 
with Gaussian generators}
Stable L\'{e}vy generators have asymmetry in the distribution, 
which also affects the mean and median.
Therefore, it is necessary to use such 
asymmetric generators for the error evaluation.
To check this, we also evaluate 
the samples in the twin experiment using the probability distribution 
generated by a cascade model with Gaussian generators
for comparison.
We take samples created by the cascade model 
with stable L\'{e}vy generators (case {\tt G3})
or Gaussian generators (case {\tt G1}), 
and then estimate the CI
via the joint PDF based on the cascade model with
the best-fitted Gaussian generators.
The results are shown in 
Fig.\,\ref{est_gauss}a, where
a significant portion ($9891$ trials out of $30000$)
of the true energy input rate (black) 
protrudes above the CI (purple).
This means that
the values of the energy input rate are underestimated if we assume 
a Gaussian distribution for the generators, and it also 
illustrates that it is inappropriate to use the statistics from
the simulations of a cascade model with Gaussian generators for 
error evaluation.
\paragraph*{Error estimation using the bootstrap method}
The simplest method for error evaluation is 
applying the bootstrap method to each profile.
However, the errors cannot be properly assessed by such 
a conventional method.
To verify this, we estimate the CIs
by applying the bootstrapping method to the twin experiment.
For each trial, we use a $1000$-member ensemble for the bootstrapping.
Each member is constructed as follows:
If a profile at the horizontal point $\vv{x}_i$
has a set of $J_i$ observations of 
the energy dissipation rate, 
$\epsilon_{r_0}(\vv{x},z^i_j),\quad j=1,2,\cdots,J_i$,
we randomly take $J_i$ samples with replacement from the set.
The results of the error evaluation of the mean dissipation rate are 
shown in Fig.\,\ref{est_gauss}b.
The CIs are evaluated very narrowly, 
and in many trials 
($22164$ trials out of $30000$), the true energy input rate (black) is
outside the CI,
which indicates that the error estimate is far too optimistic.
This illustrates that it is 
irrelevant to use the conventional
 bootstrap method for the error evaluation.

\paragraph{Comparison of skill}
For a fair comparison of skill in the above control experiments,
we examine the errors for an estimator of the energy input rate.
First, we define the estimator, $\Theta$, from the median estimate, $\gamma_{(0.5)}$, as
\begin{align}
\Theta[m] &\define a \exp{\left(\gamma_{(0.5)}[m]\right)},\\
\frac1a &= \frac1{M'} \sum_{m=1}^{M'} 
\frac{\exp{\left(\gamma_{(0.5)}[m]\right)}}{\overline{\epsilon}[m]}
,\label{prefac}
\end{align}
where $[m]$ represents the $m$-th sample, and 
$\gamma_{(0.5)}[m]$ is the median of the samples 
generated from the energy input rate $\overline{\epsilon}[m]$.
Here, $a$ is defined to empirically satisfy the unbiasedness:
\begin{align}
\frac1{M'} \sum_{m=1}^{M'} 
\frac{\Theta[m]}{\overline{\epsilon}[m]}&=1.
\end{align}
Then, in terms of $\Theta$, we define the relative error with sample size ${M'}$ as
\begin{align}
\mathrm{ind}_{M'} &\define \left(\frac1{M'} \sum_{m=1}^{M'} 
\left|\frac{\overline{\epsilon}[m]-\Theta[m]}
{\overline{\epsilon}[m]}\right|^2\right)^{1/2}.\label{errorM}
\end{align}
Note that the error might not necessarily converge when ${M'}\to \infty$ because 
the distribution of $\Theta[m]/\overline{\epsilon}[m]$
is not Gaussian.
Nevertheless, we can still evaluate the error for a finite ${M'}$ and use it for
the comparison of skill.

The errors $\mathrm{ind}_{M'}$ with ${M'}=30000$
for various conditions are listed in table\,\ref{skill2}.
The combinations of the cascade model with the L\'{e}vy generator
or Gaussian generator and
the use of joint PDF $q_3$ or marginal PDF $q_1$ are compared;
these correspond to cases {\tt L3}, {\tt L1}, {\tt G3}, and {\tt G1} above.
The error for the arithmetic mean $\widehat{\gamma}$ is also shown for reference.
The smallest error among these is for the estimator using the result of
joint PDF $q_3$ created by the cascade model with a L\'{e}vy generator ({\tt L3}).
The errors for cases {\tt L1} and {\tt G1} are comparable to that for the arithmetic mean
because these estimators are constructed from the statistics of the arithmetic mean.
This result
clearly shows that the proposed method ({\tt L3}) can be used to define an
estimator that yields superior estimates of the energy input rate
than the arithmetic mean or other methods ({\tt L1}, {\tt G3}, or {\tt G1}).

\begin{table}[ht]
\begin{center}
\caption{Comparison of the error, $\mathrm{ind}_{M'}~({M'}=30000)$, for various estimators in identical twin experiment.
L\'{e}vy/Gaussian indicates the generator used in cascade model simulation. Here,
$q_3$/$q_1$ indicates whether joint PDF or marginal PDF is used 
as the density generated by the cascade model simulation;
$\widehat{\gamma}$ indicates the arithmetic mean.
Each number in parentheses is the prefactor $a$ for the corresponding estimator. 
\label{skill2}}
\begin{tabular}{l|ll|ll}
$\mathrm{ind}_{M'}$&\multicolumn{2}{c|}{$q_3$}&\multicolumn{2}{c}{$q_1$}\\
\hline
\hline
L\'{e}vy &{\tt L3}& 0.89 {\scriptsize (0.745)}&{\tt L1} & 0.98 {\scriptsize (0.718)}\\
\hline
Gauss    &{\tt G3} & 1.0 {\scriptsize (1.61)} &{\tt G1} & 0.98 {\scriptsize (0.689)}\\
\hline
$\widehat{\gamma}$& \multicolumn{4}{c}{0.99 {\scriptsize (1.00)}}\\
\end{tabular}
\end{center}
\end{table}
\subsection*{Real data experiment}\label{sec_real}
\begin{figure}
\begin{flushleft}\hspace{4em} a) \hspace{26em} b) \end{flushleft}
    \begin{center}
\includegraphics[width=0.49\columnwidth]{est3d_arith.eps}
\includegraphics[width=0.49\columnwidth]{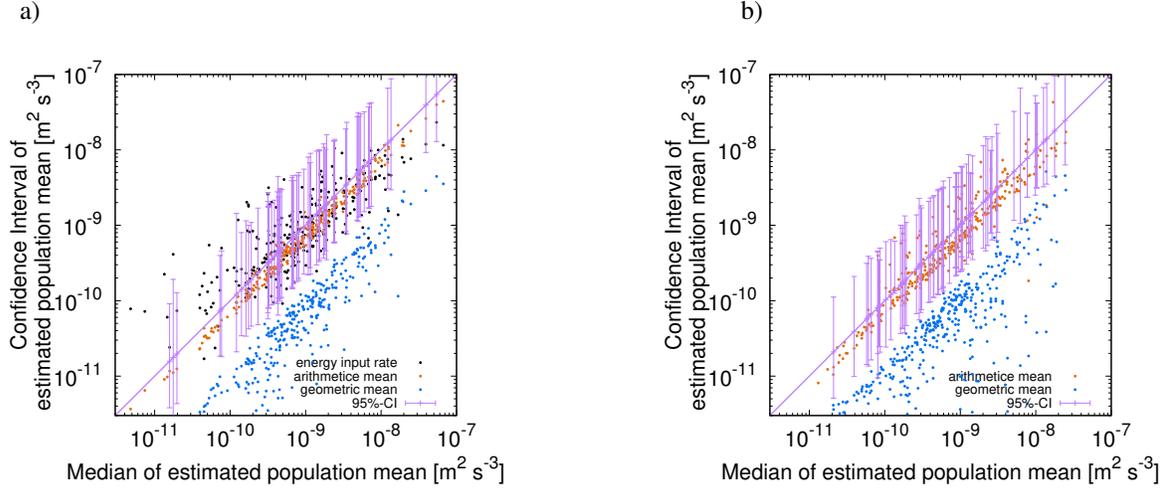}
\caption{Results of the real data experiment
compared with those of the identical twin experiment.
Median (horizontal axis) versus confidence interval (vertical axis)
of $\overline{\epsilon}$.
a) Results of identical twin experiment.
b) Results of real data experiment.
For given values of the median on the horizontal axis,
the points on the vertical axis indicate the values of 
the confidence interval (purple segment), arithmetic mean (orange), geometric mean (blue),
and energy input rate (black). 
In b, only $70$ confidence intervals out of $353$ trials  
are shown for readability.
\label{est_arith}}
    \end{center}
  \end{figure}

\begin{figure}
\begin{flushleft}\hspace{4em} a) \hspace{26em} b) \end{flushleft}
    \begin{center}
     \includegraphics[width=0.49\columnwidth]{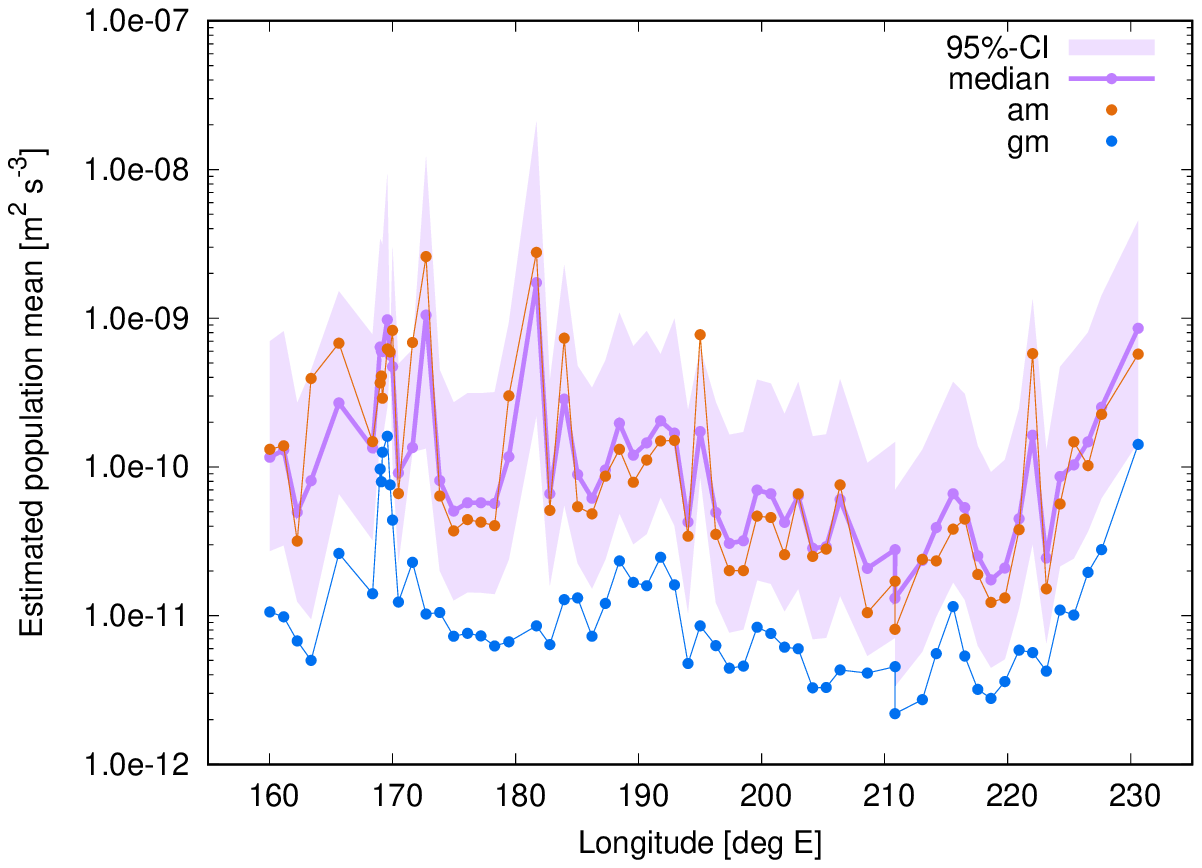}
     \includegraphics[width=0.49\columnwidth]{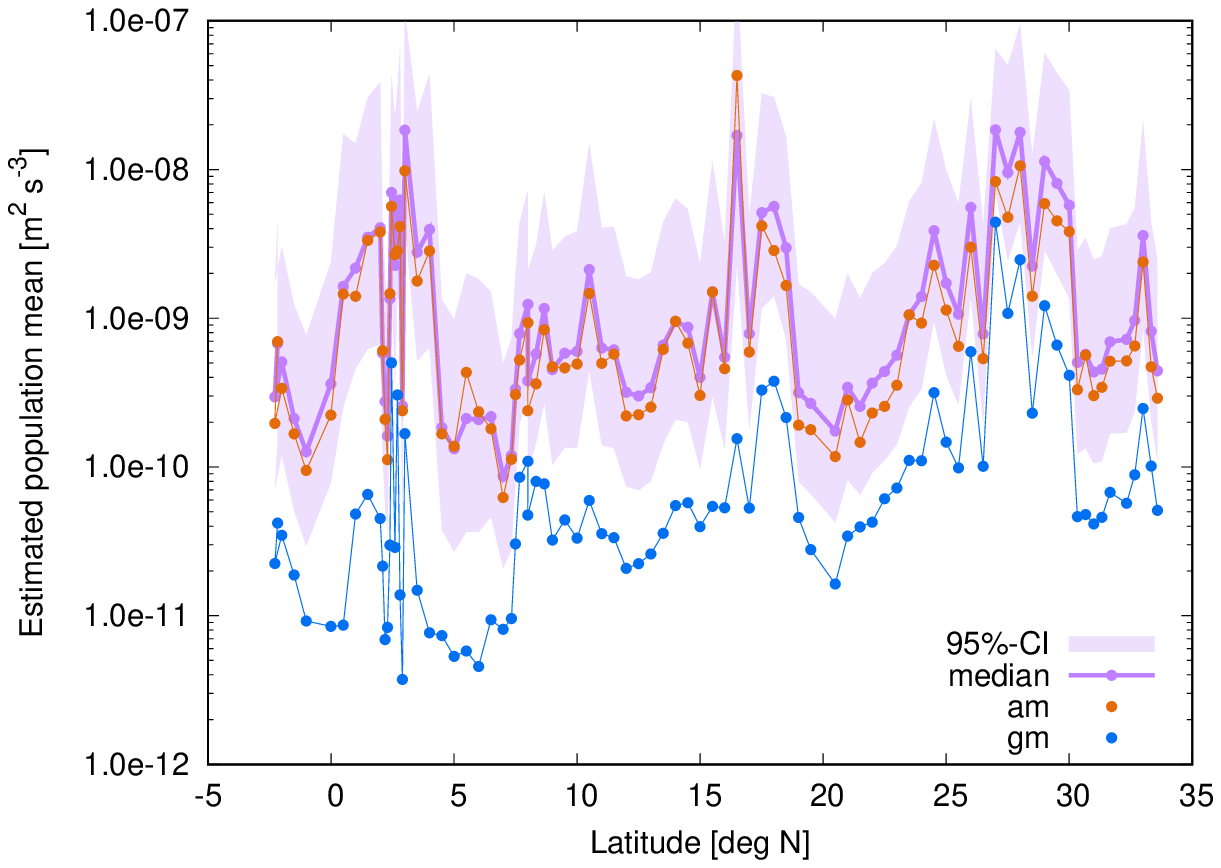}
\caption{Geographical distribution of 
median (purple dots) and confidence interval (purple shade) 
of $\overline{\epsilon}$ along (a)
$47\degree \mathrm{N}$  and (b) $137\degree \mathrm{E}$ .
The horizontal axis shows the location.
Arithmetic mean (orange) and geometric mean (blue)
are also shown.
\label{ave3d}}
    \end{center}
  \end{figure}

We applied the same procedure
as in the identical twin experiment 
to the real observational profiles
of the energy dissipation rate.
Each profile was characterised by 
$\widehat{\gamma}$, $\widehat{\gamma}-\widetilde{\gamma}$,
and $\widehat{\gamma}-\gamma^{\sharp}$, which were utilised 
as observational constraints.
By means of inversion, we derived the CI of $\overline{\gamma}$ 
for each profile at different horizontal locations.
The estimated CIs for real data are shown
in Fig.\,\ref{est_arith}b; these CIs exhibit a similar appearance 
to the ones for the identical twin experiment in Fig.\,\ref{est_arith}a.
Figure\,\ref{ave3d} shows
the estimate for CIs 
on the sections along 
$47\degree \mathrm{N}$ and $137\degree \mathrm{E}$.
Along $47\degree \mathrm{N}$, 
the median of $\overline{\epsilon}$ 
rarely exceeds $10^{-9} \unit{m^2s^{-3}}$,
except around $172\degree \mathrm{E}$, 
$180\degree \mathrm{E}$, or $50\degree \mathrm{W}$.
The peak of the arithmetic mean at $172\degree \mathrm{E}$
is approximately $2.5$ times larger than the median estimate,
which can lead to overestimation.

Along $137\degree \mathrm{E}$, 
the median shows several significant peaks
over $10^{-8} \unit{m^2s^{-3}}$
at around $2\degree \mathrm{N}$, $16\degree \mathrm{N}$,
and $27$ to $29 \degree \mathrm{N}$.
We could have underestimated
the peaks at around 
$2\degree \mathrm{N}$ and $27$ to $29 \degree \mathrm{N}$,
but overestimated the one at around 
$16\degree \mathrm{N}$,
if only the arithmetic means were used.

For the analysis of the observations 
on the section along $165\degree \mathrm{E}$,
we should consider the effects of repeated observation.
In fact, the observations 
were performed twice at some horizontal locations.
For such cases, 
we simply assume that two independent realisations 
of a common $\overline{\gamma}$ are observed. 
In this regard, the inversion formula in Eq.\,(\ref{est1i})
is modified as follows:
\begin{align}
 P(\overline{\gamma}|\widehat{\gamma}_1,\widehat{\gamma}_2)
&=
\frac{P(\widehat{\gamma}_1,\widehat{\gamma}_2|\overline{\gamma})P(\overline{\gamma})}
{\int P(\widehat{\gamma}_1,\widehat{\gamma}_2|\overline{\gamma})P(\overline{\gamma})\d \overline{\gamma}}
=
\frac{P(\widehat{\gamma}_1|\overline{\gamma})
P(\widehat{\gamma}_2|\overline{\gamma})P(\overline{\gamma})}
{\int P(\widehat{\gamma}_1|\overline{\gamma})
P(\widehat{\gamma}_2|\overline{\gamma})
P(\overline{\gamma})\d \overline{\gamma}}
=
\frac{q_1(\overline{\gamma}-\widehat{\gamma}_1)
q_1(\overline{\gamma}-\widehat{\gamma}_2)}
{\int q_1(\overline{\gamma}-\widehat{\gamma}_1)
q_1(\overline{\gamma}-\widehat{\gamma}_2)
\d \overline{\gamma}},\label{est_dble}
\end{align}
where two observations are distinguished by the subscripts $1, 2$.
This distribution is the normalised product of 
the two distributions.
When considering $\widetilde{\gamma}$ and $\gamma^{\sharp}$, 
the same procedure as in Eq.\,(\ref{est_dble}) 
is applied to $q_3(\cdot|u,v)$ instead of $q_1(\cdot)$:
\begin{align}
&P(\overline{\gamma}|\widehat{\gamma}_1,\widehat{\gamma}_2,
\widehat{\gamma}_1-\widetilde{\gamma}_1=u_1,
\widehat{\gamma}_1-\gamma^{\sharp}_1=v_1,
\widehat{\gamma}_2-\widetilde{\gamma}_2=u_2,
\widehat{\gamma}_2-\gamma^{\sharp}_2=v_2
)\nonumber \\
&=
\frac{q_3(\overline{\gamma}-\widehat{\gamma}_1|u_1,v_1)
q_3(\overline{\gamma}-\widehat{\gamma}_2|u_2,v_2)}
{\int q_3(\overline{\gamma}-\widehat{\gamma}_1|u_1,v_1)
q_3(\overline{\gamma}-\widehat{\gamma}_2|u_2,v_2)
\d \overline{\gamma}}.\label{cond_dble}
\end{align}

\begin{figure}
\begin{flushleft}\hspace{4em} a) \hspace{26em} b) \end{flushleft}
    \begin{center}
     \includegraphics[width=0.49\columnwidth]{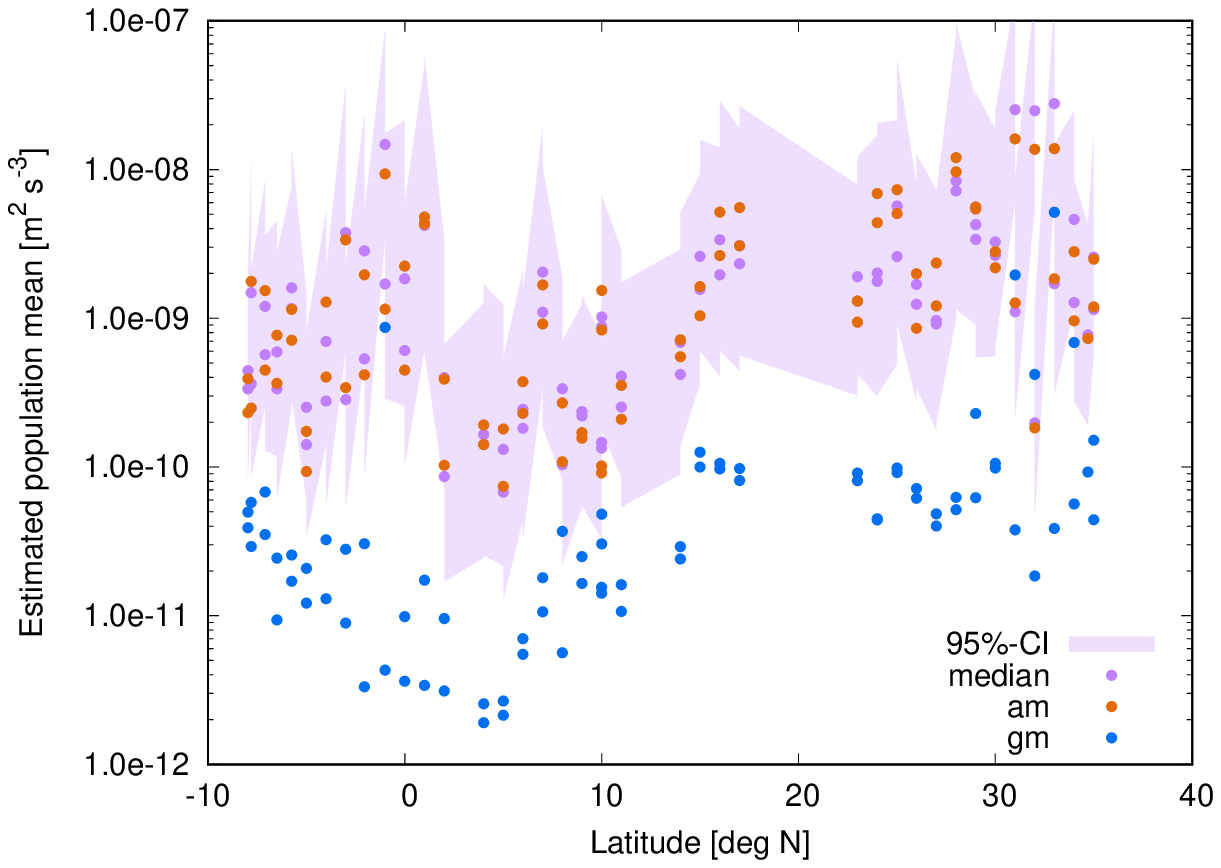}
     \includegraphics[width=0.49\columnwidth]{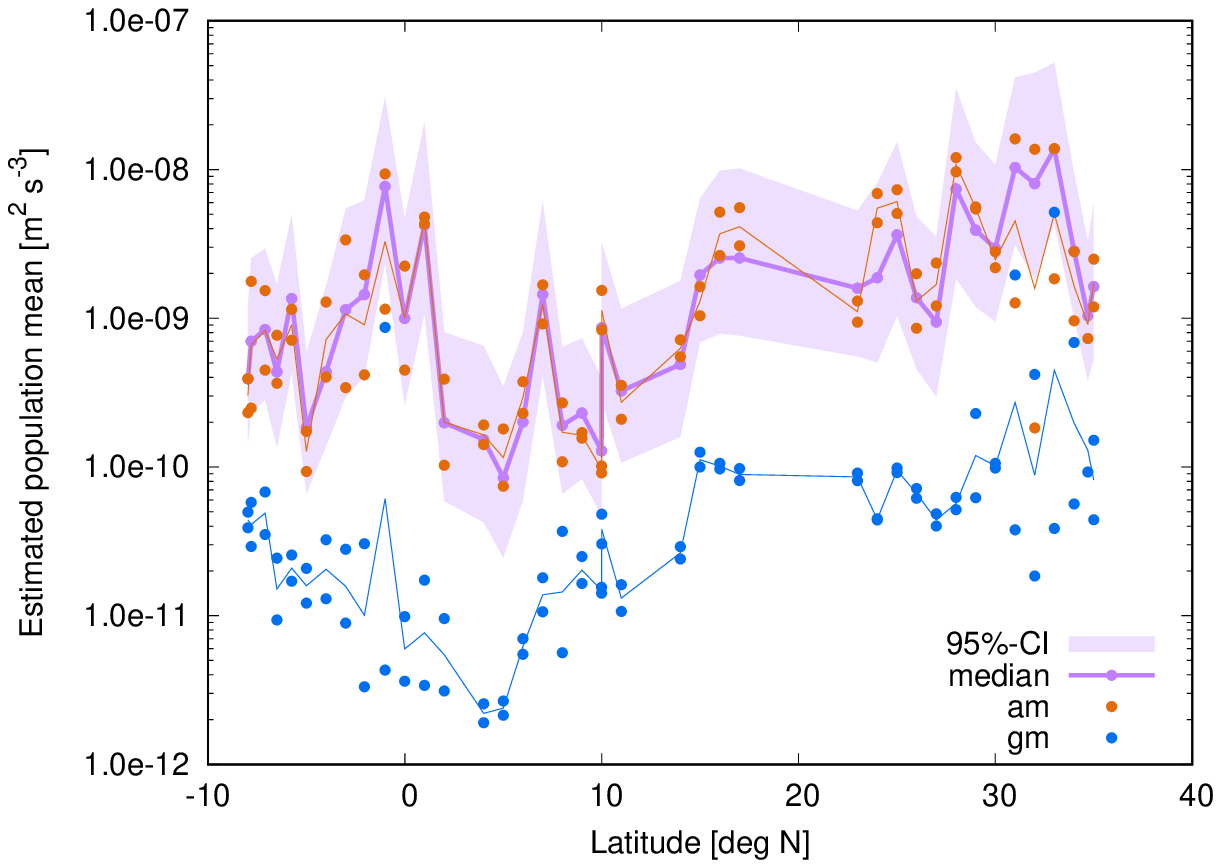}
\caption{Geographical distribution of 
median (purple dots) and confidence interval (purple shade) 
of $\overline{\epsilon}$ along $165\degree \mathrm{E}$.
The effect of repeated observation is considered in b), 
but not in a).
The horizontal axis shows the location.
Arithmetic mean (orange) and geometric mean (blue)
are also shown.
\label{ave3d2}}
    \end{center}
\end{figure}

Figure\,\ref{ave3d2}b shows the estimation of CI along $165\degree \mathrm{E}$
by taking into account the effect of repeated observation.
For comparison, the result without considering the repeated observation
is shown in Fig.\,\ref{ave3d2}a, 
where each observation is assumed 
to correspond to independent $\overline{\gamma}$.
We can see that 
the CIs become narrower 
when considering the effect of repeated 
observation.
Furthermore, along $165\degree \mathrm{E}$, 
the median shows a significant plateau 
on the order of $10^{-8} \unit{m^2s^{-3}}$
at around $30\degree \mathrm{N}$, 
and a significant peak
on the order of $10^{-8} \unit{m^2s^{-3}}$
at around $2\degree \mathrm{S}$. 

\section*{Discussion}\label{conc}
  We have analysed the observed data obtained from oceanic turbulence measurements and shown that the vertical sequences of 
energy dissipation rates in a profile have an intermittent structure that obeys a scaling law. In this study, we have laid greater emphasis on the 'energy input rate', which refers to the population mean for a profile, than on  the 'mean energy dissipation rate', which is the sample (arithmetic) mean over a profile.
Based on the scaling property, we have proposed a method of estimating the energy input rate, given the sample statistics of an observed profile.
\begin{enumerate}
\item
  For scaling within the observed profiles, the statistical properties of our data are consistent with the universal multifractal model, which has a moment scaling exponent of 
$K(q)=\left(C_1/(\alpha-1)\right)\left(q^{\alpha}-q\right)$
with a multifractal index 
$\alpha = 1.62 \pm 0.03$ and codimension of the mean $C_1 = 0.352 \pm 0.009$.
This result elucidates the universality that is inherent in the vertical structure of oceanic turbulence data.
\item 
  The energy input rate and its uncertainty can be estimated using the results of Monte Carlo simulation of the cascade model with stable L\'{e}vy generators.
This method computes the conditional probability, given the observed values of the arithmetic mean, geometric mean, and quadratic mean over a profile. The estimate provides additional information on the uncertainty of the energy input rate.
\item
Furthermore, a comparison to control experiments 
has demonstrated that the proposed method is superior to 
a simple bootstrap method, 
an inversion based on the PDF
generated by a multiplicative cascade with 
Gaussian generators, or 
an inversion based on 
the PDF of the arithmetic mean.
\item
A real data experiment using the observed profiles 
has demonstrated a geographical distribution
of the median estimates and confidence intervals of energy input rate,
providing information on the range of values in which the turbulent energy can be dissipated per unit depth at each horizontal location.
\item 
  Thus, we have found an answer to the question: 
  `How can one estimate energy input rate from the vertical profile data of the energy dissipation rate?' By analysing the intermittency in the observed data, we can construct a multiplicative cascade model based on the universal multifractal formalism that can reproduce the statistics of the data. Then, based on the observed data, the energy input rate can be estimated by inverting the probability distribution obtained from Monte Carlo simulations of the cascade model.
\item 
Since the observed sequence of 
the energy dissipation rate fluctuates greatly
due to intermittency, it is difficult to extract robust information 
from an observed profile only by examining the average.
Using the proposed method, 
it is possible to estimate the population mean for a profile 
even when repeated observations cannot be made at the same horizontal position.
In other words, we can distinguish, to some extent,
whether a profile shows an occasional large mean or whether 
the population mean itself is large. 
Therefore, more information can be extracted from 
a small amount of turbulence observation data, 
which can be a great advantage in regional data analysis.
\item
Theoretically, this technique can easily be extended by utilising more statistics over a profile besides the arithmetic mean, geometric mean, or quadratic mean. Note however that such an extension may easily suffer from the curse of dimensionality, and thus, it can become impractical.
\item 
  Even though we have used a discrete cascade model for simplicity and computational viability, we can extend it to a continuous cascade 
  \cite{schmitt2001stochastic}, which may improve the estimation accuracy at the cost of increased computational burden.
\item To investigate the scaling of the velocity spectrum in the horizontal and vertical directions, we should apply anisotropic scaling theory 
(e.g., the Kolmogorov-Bolgiano-Obukhov model\cite{lovejoy2013weather}).
However, this study only aimed at investigating the scaling of the energy dissipation rate in the vertical direction in a purely statistical manner.
Nevertheless, because the energy dissipation rate is one of the key quantities in scaling analysis, our results will contribute to further studies on how intermittency affects various scaling behaviours of
turbulence in buoyancy-driven stratified fluids.
\end{enumerate}

\bibliography{multifractal}

\section*{Acknowledgements}
  Helpful comments that were received from Yutaka Yoshikawa (Kyoto University) are highly appreciated.
  This work was partially supported by a Grant-in-Aid for Scientific Research on Innovative Areas (MEXT KAKENHI-JP15H05817/JP15H05819).
We also thank the members of the project for their valuable discussions on the concept and methodology. We would like to thank Editage for English language editing. All numerical simulations were performed on the JAMSTEC Data Analyzer (DA) system. The ocean turbulence dataset is under preparation for public release by the Atmosphere and Ocean Research Institute, University of Tokyo.

\section*{Author contributions}
  SM and SO proposed the main problem. IY compiled the observational data. IY, SK, SM, and SO helped formulate the hypothesis. NS proposed the method, performed the statistical analyses, and prepared the manuscript with contributions from all co-authors.

\section*{Competing interests}
The authors declare no competing interests.

\section*{Supplementary information}
Refer to the supplementary information file.
\end{document}